\documentclass[showpacs,twocolumn,superscriptaddress]{revtex4-2}

\usepackage{graphicx}
\usepackage{physics}
\usepackage{wrapfig}
\usepackage{changepage}
\usepackage{amsmath,amssymb,amsthm,mathrsfs,amsfonts,dsfont,mathtools}
\usepackage{algorithm}
\usepackage{algpseudocode}
\usepackage[caption=false]{subfig}
\usepackage{epsfig}
\usepackage{color,xcolor}
\usepackage{adjustbox}
\usepackage{tabularx}
\usepackage{array}
\usepackage{multirow}
\usepackage{tabu}
\usepackage{bm}
\usepackage{enumerate}
\usepackage{hyperref}
\usepackage{newtxtext,newtxmath}
\usepackage{makecell} 
\usepackage{url}
\usepackage{booktabs}
\usepackage{nicefrac} 
\usepackage{makeidx}  
\usepackage{leftindex}

\usepackage{cleveref}
\usepackage{mwe}

\usepackage{comment}
\usepackage{xurl}

\begin{document}
\title{Hamiltonian formulations of centroid-based clustering}

\author{Myeonghwan Seong}
\email{europa0306@yonsei.ac.kr}
\affiliation{Department of Statistics and Data Science, Yonsei University, Seoul 03722, Republic of Korea}
\author{Daniel K. Park}
\email{dkd.park@yonsei.ac.kr}
\affiliation{Department of Statistics and Data Science, Yonsei University, Seoul 03722, Republic of Korea}
\affiliation{Department of Applied Statistics, Yonsei University, Seoul 03722, Republic of Korea}

\begin{abstract}
    Clustering is a fundamental task in data science that aims to group data based on their similarities. However, defining similarity is often ambiguous, making it challenging to determine the most appropriate objective function for a given dataset. Traditional clustering methods, such as the $k$-means algorithm and weighted maximum $k$-cut, focus on specific objectives---typically relying on average or pairwise characteristics of the data---leading to performance that is highly data-dependent. Moreover, incorporating practical constraints into clustering objectives is not straightforward, and these problems are known to be NP-hard. In this study, we formulate the clustering problem as a search for the ground state of a Hamiltonian, providing greater flexibility in defining clustering objectives and incorporating constraints. This approach enables the application of various quantum simulation techniques, including both circuit-based quantum computation and quantum annealing, thereby opening a path toward quantum advantage in solving clustering problems. We propose various Hamiltonians to accommodate different clustering objectives, including the ability to combine multiple objectives and incorporate constraints. We evaluate the clustering performance through numerical simulations and implementations on the D-Wave quantum annealer. The results demonstrate the broad applicability of our approach to a variety of clustering problems on current quantum devices. Furthermore, we find that Hamiltonians designed for specific clustering objectives and constraints impose different requirements for qubit connectivity, indicating that certain clustering tasks are better suited to specific quantum hardware. Our experimental results highlight this by identifying the Hamiltonian that optimally utilizes the physical qubits available in the D-Wave System. 
\end{abstract}

\maketitle
\def\one{{\mathchoice {\rm 1\mskip-4mu l} {\rm 1\mskip-4mu l} {\rm\mskip-4.5mu l} {\rm 1\mskip-5mu l}}}

\section{Introduction}
Quantum machine learning (QML) offers new possibilities and approaches to address various challenges in data science, pushing the boundaries of existing methods. Among its potential applications, clustering is a widely used technique in numerous domains of pattern recognition and data mining, such as image recognition, social network analysis, customer segmentation, and anomaly detection ~\cite{coleman1979image,baraldi1999survey,jain1999data,madeira2004biclustering,wu2005research,handcock2007model,agrawal2015survey,saxena2017review}. In addition, clustering has found increasing applications in drug discovery, aiding in the selection of potential leads, mapping protein binding sites, and designing targeted therapies ~\cite{voicu2020rcdk,dara2022machine,mak2023artificial}.

Despite its broad applicability and importance, clustering encounters several challenges from an optimization perspective~\cite{jain2010data,xu2015comprehensive,ezugwu2022comprehensive}. A primary issue is the ambiguity in defining the objective function for clustering. As there is no ground truth, it is often unclear which criteria should be used to group the target dataset, requiring the analyst to make subjective decisions about what constitutes similarity. A common approach involves using distance measures to quantify similarity. 
However, this approach still requires determining whether to rely on local information, such as the pairwise distance between individual data points, or global information, such as the distance between a data point and the centroid of a cluster. When using local information, clustering can be formulated as combinatorial optimization and approached by solving the maximum $k$-cut problem, which corresponds to maximizing dissimilarity between clusters. On the other hand, an example of using global information is the $k$-means clustering algorithm, which is equivalent to minimizing the variance within clusters by focusing on the distance to centroids. Yet, the challenge remains in how to effectively incorporate both or potentially other objectives for improved clustering. Another critical challenge is that even if the analyst decides to use either local or global information as described in constructing the objective function, finding the global solution is intractable. This intractability arises because the cardinality of the feasible set (i.e., the number of clustering configurations) grows exponentially with the number of data points and due to the non-convexity of the optimization landscape. 
As a result, in practice, polynomial-time approximate algorithms are employed to obtain good local solutions. This highlights the need for developing more efficient optimization algorithms that can either improve solution quality, reduce runtime, or accomplish both. In addition, existing polynomial-time algorithms often require a random initial cluster assignment, and both the quality of the solution and the convergence speed can be highly sensitive to this choice of initialization.
\begin{figure}[!h]
    \begin{center}
    \includegraphics[width=1\columnwidth]{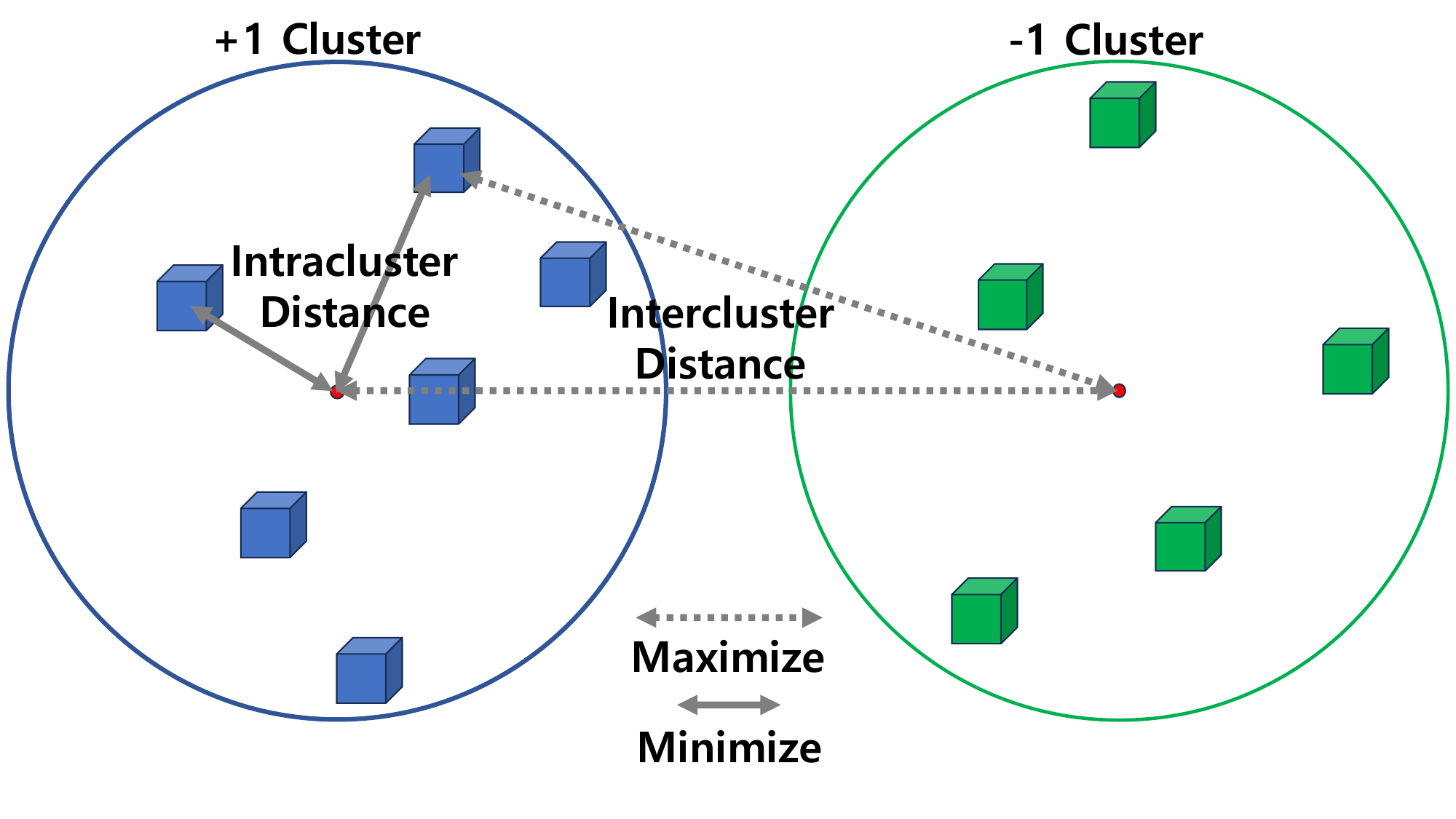}    
    \end{center}
    \caption{Illustration of the intracluster distance and the intercluster distance. Unlike supervised learning, unsupervised learning cannot utilize a loss function with exact labels. In a clustering approach, the loss function can be created based on how well the hypothesis separates data points into their appropriate groups. We customized and combined intracluster distance, which measures how tightly data points are clustered together within a cluster, and intercluster distance, which measures how far apart different cluster centers are, as weighted criteria in QUBO formula. By optimizing this loss function, we were able to solve the clustering problem.}\label{fig:1}
\end{figure}

To address these challenges, we develop a unified framework that incorporates multiple data characteristics---such as local and global information---into the optimization objective, with the flexibility to assign arbitrary weights to specify their relative importance in clustering. Our approach begins by decomposing the problem of finding $k$ clusters into hierarchical clustering, where each level of the hierarchy consists of binary clustering. We then introduce a method to incorporate centroids as variables within the objective function of a combinatorial optimization problem. This formulation enables various centroid-based binary clustering models, such as those that account for intercluster distance, intracluster distance, or both (see Fig.~\ref{fig:1}), to be cast as combinatorial optimization problems. Furthermore, the objective function with centroid variables can be linearly combined with that of the weighted max-cut problem into a single, unified objective. Solving an unconstrained combinatorial optimization problem with binary variables can be mapped to the problem of finding the ground state (i.e., the eigenstate with the lowest eigenvalue) of a spin Hamiltonian~\cite{10.3389/fphy.2014.00005}. This mapping offers a crucial benefit: it enables the problem to be solved on a quantum computer using quantum simulation techniques such as those based on quantum phase estimation and amplitude amplification~\cite{dong2022ground, poulin2009preparing, ge2019faster, lin2020near, zeng2023universal}, the variational quantum eigensolver~\cite{peruzzo_variational_2014,McClean_2016,cerezo2020variational}, the quantum approximate optimization algorithm (QAOA)~\cite{farhi2014quantum}, quantum annealing~\cite{johnson2011quantum}, or quantum-inspired algorithms~\cite{jiang2023efficient}.
Notably, QAOA and quantum annealing do not require a random initial cluster assignment, as their initial quantum state is a uniform superposition of all computational basis states. This means that the algorithms begin with all possible clustering configurations, each assigned equal weight. Consequently, these approaches are free from the sensitivity to initial conditions, unlike classical polynomial-time algorithms. Moreover, formulating clustering as a combinatorial optimization problem is advantageous when incorporating constraints, as constraints can also be formulated as combinatorial optimization problems and included as penalty terms in the objective function.

We benchmark the effectiveness of our approach and the proposed Hamiltonian formulations (i.e. the combinatorial optimization problems) using several datasets: Iris, Wine, a subset of MNIST, and a synthetic Gaussian overlapping dataset. To evaluate performance, we employed the Silhouette Score (SS)~\cite{rousseeuw1987silhouettes} and the Rand Index (RI)~\cite{rand1971objective} as metrics, conducting comparative analysis with the $k$-means algorithm~\cite{pedregosa2011scikit} and the weighted max cut. Initially, we assessed the performance of each Hamiltonian by searching for its exact solutions using a brute-force algorithm, in order to establish a benchmark for comparing theoretical predictions and practical outcomes. We then empirically tested the efficacy of our Hamiltonian formulations using simulated annealing and quantum annealing on the D-Wave Advantage System 6.4~\cite{dwave6.4}. Furthermore, we expanded our investigation to constrained clustering, incorporating Must-Link (ML), Cannot-Link (CL), and cluster size constraints. By doing so, we highlight the advantages of our centroid-based method, focusing on its ability to manage complex data structures and accommodate real-world data constraints. The following sections will elaborate on our methodological framework, the tailored Hamiltonian designs for clustering objectives, and the results of our comparative analysis, emphasizing the adaptability of our approach.

\section{Related work}
\label{sec:related}

In this section, we review prior research efforts that framed centroid-based clustering as a combinatorial optimization problem. Several studies have approached this with a particular focus on Quadratic Unconstrained Binary Optimization (QUBO) formulations.
Ref.~\cite{bauckhage2018ising} and Ref.~\cite{arthur2021balanced} explored the representation of cluster centroids in QUBO under the assumption of equal cluster sizes, typically in the context of $k$-means clustering. Ref.~\cite{bauckhage2019qubo} extended this work by introducing a QUBO formulation of the $k$-medoids approach, which differs from $k$-means clustering by selecting $k$ representative data points (medoids) as cluster centers instead of calculating centroids based on the mean of the data points.

These works illustrate that centroid-based clustering algorithms, like $k$-means and $k$-medoids, can be formulated as combinatorial optimization problems, specifically QUBO, albeit under restricted conditions. A recurring assumption in these studies is the uniform distribution of data, which undesirably constrains clusters to be of approximately equal size. Moreover, the use of synthetic data in experiments raises concerns about the generalizability of these methods to real-world data. While Ref.~\cite{matsumoto2022distance} proposed an iterative fractional cost approach to address the issue of uneven data distributions, their solution significantly increases computational complexity due to the need for hyperparameter tuning and iterative recalculations.

In contrast to previous approaches, our method does not require predefined cluster sizes or rely on computationally intensive iterative processes. It directly incorporates the number of data points in each cluster as a variable in the objective function, eliminating the assumption of fixed cluster sizes. This enables greater flexibility and adaptability to a wider range of data distributions.

\section{Methodology}
We begin by discussing the process of mapping clustering problems to combinatorial optimization problems. To represent the assignment of $N$ data points into two clusters, we use a binary variable $z \in \{ -1, +1 \}^N$. Each element $z_i$ indicates the cluster assignment of the $i$th data point, $x_i$, where $x_i\in\mathbb{R}^{d}$ is the representation of the data in the feature space. This feature space representation is obtained after applying any necessary pre-processing techniques, such as Principal Component Analysis (PCA), normalization, or standardization.
 
In the Hamiltonian formulation, we introduce $z_i' = (1+z_i)/2\in \{0,1\}$ to denote the computational basis state of the $i$th qubit, representing the cluster assignment. The variables $z$ and $z'$ variables correspond to the eigenvalues and the eigenstates of the Pauli $Z$ operator, respectively: $Z|0\rangle = +|0\rangle$ and $Z|1\rangle = -|1\rangle$.
In general, the objective function subject to minimization in QUBO problems can be expressed as
\begin{equation} \label{eq:qubo}
    f(z) = a_0 + \sum_{i<j} a_{ij}z_iz_j + \sum_{i=1}^Na_iz_i.
\end{equation}
In the context of clustering, $a_0$ is a constant independent of $z$, $a_{ij}$ represents the relationship between data points $x_i$ and $x_j$, $a_i$ reflects characteristics of each individual data point. The combinatorial optimization problem can be mapped to finding the smallest eigenvalue and the corresponding eigenvector of the Hamiltonian for a finite-dimensional quantum system. The corresponding Hamiltonian is obtained by replacing $z_i$ with the Pauli $Z$ operator and $1$ with the identity operator acting on the $i$th qubit:
\begin{align} 
    H &= \sum_{z\in\lbrace -1,+1\rbrace^N}f(z)|z'\rangle\langle z'| \\&= a_0 I +  \sum_{i<j}a_{ij}Z_iZ_j + \sum_{i=1}^N a_i Z_i.\label{eq:hamiltonian}
\end{align}
where $z'\in \lbrace 0,1\rbrace^N$ is obtained by mapping every element of $z$ as $(1+z_i)/2$, and $I$ is the identity matrix.

Existing QUBO-based clustering algorithms typically rely solely on pairwise distances between data points. In this case, $a_0=0$ and $a_i=0$ for all $i$, leading to an optimization problem of the form
\begin{equation}
    \label{eq:maxcut}
    \min_{z \in \{-1,+1\}^N} \sum_{i<j}^{N} a_{ij}z_iz_j,
\end{equation}
where $a_{ij}\ge 0$ represents the dissimilarity measure between the $i$th and $j$th data points. For instance, $a_{ij}=\Vert x_i - x_j\Vert_2$. Equivalently, this problem can be formulated as finding the ground state (i.e. the lowest-energy state) of the Hamiltonian,
\begin{equation}
    H = \sum_{i<j}a_{ij}Z_iZ_j.
\end{equation}
This optimization problem is also known as the weighted max-cut problem on a graph. However, this formulation of clustering neglects global information, such as centroids, in the optimization process. This limitation is primarily due to the computational complexities and challenges involved in representing centroids within the combinatorial optimization framework, unless there is prior knowledge that the dataset is evenly distributed among clusters~\cite{Kumar_2018,arthur2021balanced,date2021qubo} as noted in Section~\ref{sec:related}.

To incorporate the centroid information into the combinatorial optimization (e.g. QUBO) framework, we introduce the variables $N_{+}$ and $N_{-}$, which represent the number of data points assigned to +1 and -1, respectively, and add these variables into the objective function. The number of data points assigned to each cluster can be computed as
\begin{equation}
     N_{\pm} = \sum_{i=1}^{N}\frac{1\pm z_i}{2}.
\end{equation}
These variables serve as the building blocks for constructing the desired objective function, along with any necessary constraints. The centroids of the two clusters can then be expressed as
\begin{equation}
\label{centroid}
     \mu_{\pm} = \frac{1}{N_{\pm}}\sum_{i=1}^{N}x_i\frac{1\pm z_i}{2}.
\end{equation}
Moreover, for a given dataset $x=\lbrace x_i\rbrace_{i=1}^N$, we define the distance function $l:\mathbb{R}^d\times \lbrace +1,-1\rbrace^{N}\times \lbrace +1,-1\rbrace \rightarrow \mathbb{R}_{\ge 0}$ as
\begin{equation}
\label{eq:centroid_loss}
    l(\mu,z,s) = \sum_{i=1}^N \Vert x_i - \mu\Vert_2 ^2\frac{1+ s z_i}{2}.
\end{equation}
Here, $(1+sz_i)/2$ acts as an indicator function, taking the value 1 if $z_i =s$ (i.e., if $x_i\in C(s)$, where $C(s)$ denotes the cluster labeled by $s$), and 0 otherwise. Thus, this function computes the total distance between the centroid $\mu$ and the data points in the cluster labeled by $s$. For instance, $l(\mu_{\pm},z,\pm 1)$ measures the total distance between the centroid $\mu_{\pm}$ and the data points grouped in the cluster labeled $\pm 1$. (i.e., the total intracluster distance). On the other hand, $l(\mu_{\pm},z,\mp 1)$ calculates the total distance between the centroid of the $\pm 1$ cluster and the data points labeled the $\mp 1$ (i.e., the total intercluster distance).

To construct a clustering objective function that incorporates centroid information, one can take a linear combination of the distance functions $l(\mu,z,s)$. However, this approach poses two challenges when applying the Hamiltonian approach to solve the optimization problem. First, it is non-trivial to map the binary variables within the $1/N_{\pm}$ term into the Hamiltonian framework. Second, these denominators can cause numerical instability if all (or nearly all) data points are assigned to one of the clusters.
To address these issues, we multiply the objective function by suitable powers of $N_{\pm}$ to eliminate the denominators and prevent numerical instability. As detailed in Appendix \ref{sec:Appendix_A}, the terms involving $N_{\pm}$ in the denominators appear with powers of either 1 or 2 from when the functions are combined for optimization.
Accordingly, we apply the necessary multiplicative factors to cancel out these terms while minimizing deviations from the original clustering objectives. By linearly combining these modified functions, optimization problems that focus on minimizing intracluster distances, maximizing intercluster distances, or both can be transformed into a Hamiltonian problem.
In the following, we present three specific examples of such centroid-based objective functions. The corresponding Hamiltonians can be derived using a similar procedure as explained in Eqs~(\ref{eq:qubo}) to (\ref{eq:maxcut}), by replacing the scalar $1$ with a $2^N$-dimensional identity matrix and the binary variables $z_i$ with the Pauli $Z_i$ operators acting on the $i$th qubit. The multiplications of the binary variables $z_i$ (e.g. $z_iz_j$ or $z_iz_jz_k$ corresponds to the tensor products of the Pauli $Z_i$ operators (e.g. $Z_i\otimes Z_j$ or $Z_i\otimes Z_j \otimes Z_k$).

\subsection{Intracluster Distance}
\label{sec:intra}
We start by setting up the optimization problem aimed at minimizing intracluster distances. This is achieved by linearly combining $l(\mu_{+},z,+1)$ and $l(\mu_{-},z,-1)$ and scaling the result by an appropriate multiplicative factor, as shown below:
\begin{equation}
    \label{eq:intra}
    \min_{z \in \{-1,+1\}^N}  N_{+}^2N_{-}^2\left[l(\mu_{+},z,+1) + l(\mu_{-},z,-1)\right].    
\end{equation}
This formulation encourages clusters to concentrate around their centroids by minimizing intracluster variance, which is conceptually equivalent to the objective of the well-known $k$-means algorithm.

However, the minimum of the objective function in Eq.~\eqref{eq:intra} can be achieved by setting either $N_+$ or $N_-$ to zero, leading to a trivial solution that does not represent useful clustering. To prevent this, we multiply $l(\mu_{+},z,+1)$ by $N_{+}^2$ and $l(\mu_{-},z,-1)$ by $N_{-}^2$, focusing on the respective clusters. The problem can then be reformulated as:
\begin{align}
    \label{eq:intra2}
    \min_{z \in \{-1,+1\}^N} N_+^2 l(\mu_{+},z,+1) + N_-^2 l(\mu_{-},z,-1).
\end{align}
Notably, in each intracluster distance term, either $N_+$ or $N_-$ appears only with a power of 1 (see Appendix~\ref{sec:Appendix_A}). Thus, multiplying each term by a linear factor $N_{\pm}$ suffices to eliminate the denominator. However, using higher-order factors, such as the quadratic term $N_{\pm}^2$, not only removes the denominator but also reflects the influence of cluster sizes into the optimization process. 
To analyze how different powers of $N_{\pm}$ influence the clustering results, we conducted simulations using both $N_{\pm}^2$ and $N_{\pm}$. The results obtained by scaling with $N_{\pm}^2$ are labeled as \textit{Intra}, whereas those obtained by scaling with $N_{\pm}$ are labeled as \textit{Intra}$^*$.

\subsection{Intercluster Distance}
\label{sec:inter}
To achieve well-separated clusters, it is beneficial to consider intercluster distance, which aims to maximize the separation between different clusters. While minimizing intracluster distance enhances cohesion within each cluster, it may introduce ambiguity near adjacent clusters, especially when boundaries are unclear. By focusing on intercluster separation, we can better distinguish data points near ambiguous or overlapping boundaries, thereby improving the overall clustering performance. Following a similar approach to that used for intracluster distances, the objective function is constructed by linearly combining $l(\mu_-,z,+1)$ and $l(\mu_+,z,-1)$, with both terms multiplied by $N_{+}^2N_{-}^2$. Since each intercluster distance term includes either $1/N_+^2$ or $1/N_-^2$ (see Appendix~\ref{sec:Appendix_A}), multiplying the entire linear combination by $N_{+}^2N_{-}^2$ is necessary to cancel these denominators. The resulting optimization problem for intercluster distance is then defined as follows:
\begin{equation}
    \label{eq:inter}
    \min_{z \in \{-1,+1\}^N} - N_{+}^2N_{-}^2 \left[l(\mu_-,z,+1) + l(\mu_+,z,-1)\right].
\end{equation}
In this formulation, we maximize the squared distance of each data point to the centroid of the opposite cluster, encouraging the data points to be as far as possible from the other cluster. We observe that this approach enhances clustering performance, particularly in cases where cluster boundaries are not clearly defined (see Sec.~\ref{sec:Experiments}). 

\subsection{Combining intra and intercluster distances}
\label{sec:intrainter}
Now, we can integrate both intracluster and intercluster distances within a unified framework. By simultaneously optimizing these distances, we aim to strengthen the compactness within clusters while enhancing the separation between different clusters. This can be achieved by linearly combining Eq.~\eqref{eq:intra} and Eq.~\eqref{eq:inter}, with the multiplicative factor $N_+^2N_-^2$, which removes the denominators in both the intracluster and intercluster distance terms. The resulting optimization problem is
\begin{align}
    \label{eq:intra and inter}
    \min_{z \in \{-1,+1\}^N} N_+^2&N_-^2[l(\mu_+,z,+1) 
    + l(\mu_-,z,-1) \nonumber\\
    &- l(\mu_-,z,+1)-l(\mu_+,z,-1)].
\end{align}
The optimization aims to assign each data point $x_i$ to a cluster label $z_i \in \{-1,+1\}$ such that the overall intracluster distances are minimized while the intercluster distances are maximized. Specifically, the function promotes tight clustering by minimizing the distances between data points and the centroid of their assigned cluster. At the same time, it enhances separation by maximizing the distances between data points and the centroid of the opposite cluster. By optimizing over all possible assignments of $z_i$, we seek a clustering configuration where data points are closely grouped around their respective centroids and well-separated from the other cluster.

By rearranging Eq.~\eqref{eq:intra and inter} (see Appendix~\ref{sec:Appendix_A3}), the combined objective function can also be expressed as:
\begin{equation}
    \label{eq:intra and inter2}
    \min_{z \in \{-1,+1\}^N} -NN^2_{+}N^2_{-}\Vert\mu_{+} - \mu_{-} \Vert_2^2.
\end{equation}
This expression reveals that optimizing the combined intracluster and intercluster distances is equivalent to maximizing the squared distance between the cluster centroids. Therefore, by optimizing the combined objective function in Eq.~\eqref{eq:intra and inter}, we inherently maximize the separation between the centroids of the two clusters.

\subsection{Constrained clustering}
In practice, analysts often need to perform clustering under constraints, which are dictated by task requirements or the available information. These constraints ensure that clustering not only groups data effectively but also adheres to the underlying structure and expert knowledge specific to the domain. According to Ref.~\cite{ganccarski2020constrained}, these constraints typically fall into three main categories: labeling constraints, cluster constraints and comparison constraints.

Labeling constraints are based on preassigned labels from domain knowledge, guiding the clustering algorithm to ensure that labeled objects are assigned to the correct groups. Cluster constraints focus on the characteristics of the clusters, such as the desired number of clusters or restrictions on cluster size or density. Comparison constraints include Must-Link (ML) and Cannot-Link (CL) relations, which specify whether certain objects should or should not be placed in the same cluster based on their inherent relationships. This approach allows users to specify relationships between data points even in the absence of class labels. In our framework, these constraints can be incorporated by augmenting the objective function with penalty terms that increase the objective value when the constraints are violated.

To implement labeling constraints, we modify the objective function to penalize incorrect cluster assignments. If the $i$th data point is labeled as $+1$ (corresponding to the $z_i = 1$), we add a term $- \lambda_p z_i$ with $\lambda_p > 0$ to the objective function. Similarly, if the data point is labeled $-1$ (corresponding to the $z_i = -1$), we add $+\lambda_p z_i$. This ensures that labeled points are assigned to the correct cluster, minimizing the penalty function when the cluster assignment matches the provided labels.
For cluster constraints, if the goal is to ensure that a specific number of data points are assigned to each cluster, we can modify the objective function as
\begin{equation}
\label{eq:cardinality constrained}
    f(z) + \lambda_{p}\left(C-\sum_{i=1}^{N}z_i\right)^2.
\end{equation}
Here, $C$ represents the desired difference in the number of data points between two clusters, and $\lambda_{p}>0$ is the hyperparameter controlling this aspect. Since $z_i\in\{ -1,+1\}$, the term $\sum_{i=1}^N z_i$ evaluates the difference in the number of data points between the two clusters for a given cluster assignment. Consequently, the second term in Eq.~(\ref{eq:cardinality constrained}) becomes zero only when the constraint is satisfied, while the objective value increases quadratically with deviation from the desired cluster sizes.

Comparison constraints, such as Must-Link (ML) and Cannot-Link (CL), can also be incorporated. Using the penalty term described in Ref.~\cite{wang2010flexible}, where $Q_{ij}=+1$ for Must-Link and $Q_{ij}=-1$ for Cannot-Link, we modify the objective function as
\begin{equation}
    \label{eq:MLandCL constrained}
    f(z) - \lambda_{p} \sum_{i<j} Q_{ij} z_i z_j.
\end{equation}
The second term ensures that the objective value increases when Must-Link or Cannot-Link constraints are violated, thereby seeking a solution that satisfies these pairwise relationships.

Note that the constraints are incorporated via the penalty method, where the hyperparameter $\lambda_p$ controls the strength of constraint enforcement. Choosing an appropriate value for $\lambda_p$ is crucial, as excessively large values may overly restrict the optimization, while very small values may fail to enforce constraints effectively. Common approaches for selecting $\lambda_p$ include grid search, random search~\cite{10.5555/2188385.2188395}, and adaptive methods~\cite{NIPS2012_05311655}.

\subsection{$k$-Clustering}

\begin{algorithm}[H]
\caption{Hamiltonian $k$-clustering}
\label{algorithm1}
\begin{algorithmic}[1] 
\State $k \gets$ the desired number of clusters
\State $clusters \gets \left\lbrack\; \right\rbrack$ \Comment{Initialize an empty list of clusters}
\While{number of clusters in $clusters < k$}
    \State Construct a customized Hamiltonian
    \State Determine the ground state of the Hamiltonian using a quantum algorithm
    \State Extract the clusters from the binary solution
    \State Repeat steps 4-6 for each extracted cluster
    \State Append the new clusters to the $clusters$ list
\EndWhile
\State \Return $clusters$
\end{algorithmic}
\end{algorithm}

Building upon the work of Ref.~\cite{Kumar_2018}, we briefly discuss a $k$-clustering method inspired by hierarchical clustering techniques. To formalize this approach, we present the Hamiltonian $k$-clustering algorithm, shown in Algorithm~\ref{algorithm1}. This method iteratively performs binary clustering, eliminating the need for one-hot encoding for each cluster and avoiding complex constraint penalty terms, such as those ensuring that each data point belongs to only one cluster. Consequently, this approach simplifies the clustering process, reduces the problem size, and enhances scalability, making it more suitable for the current capabilities of quantum annealers. Furthermore, this method can provide hierarchical insights into the data structure by unveiling nested cluster relationships. Thus, Hamiltonian clustering can be extended beyond binary clustering to general clustering problems.
In the following section, we present experimental results that validate the effectiveness of our proposed method.

\section{Experiments}\label{sec:Experiments}
To assess the effectiveness of our array of customized Hamiltonians, we conducted experimental analyses using the Silhouette Score (SS) and the Rand Index (RI) as primary performance metrics, with comparisons to the $k$-means algorithm and the weighted MaxCut. The Silhouette Score evaluates cohesion within clusters and separation between clusters, while the Rand Index measures agreement with the ground truth by calculating the true positives and true negatives in the clustering results. In addition to these primary metrics, we examined other aspects of clustering performance, such as the distances between cluster centroids, intracluster distances (the sum of distances within clusters), and intercluster distances (the sum of distances between clusters). These additional results are summarized in the tables in Appendix~\ref{sec:Appendix_B}.

\subsection{Exact solutions}
\begin{figure}[t]
    \begin{center}
    \includegraphics[width=0.5\textwidth]{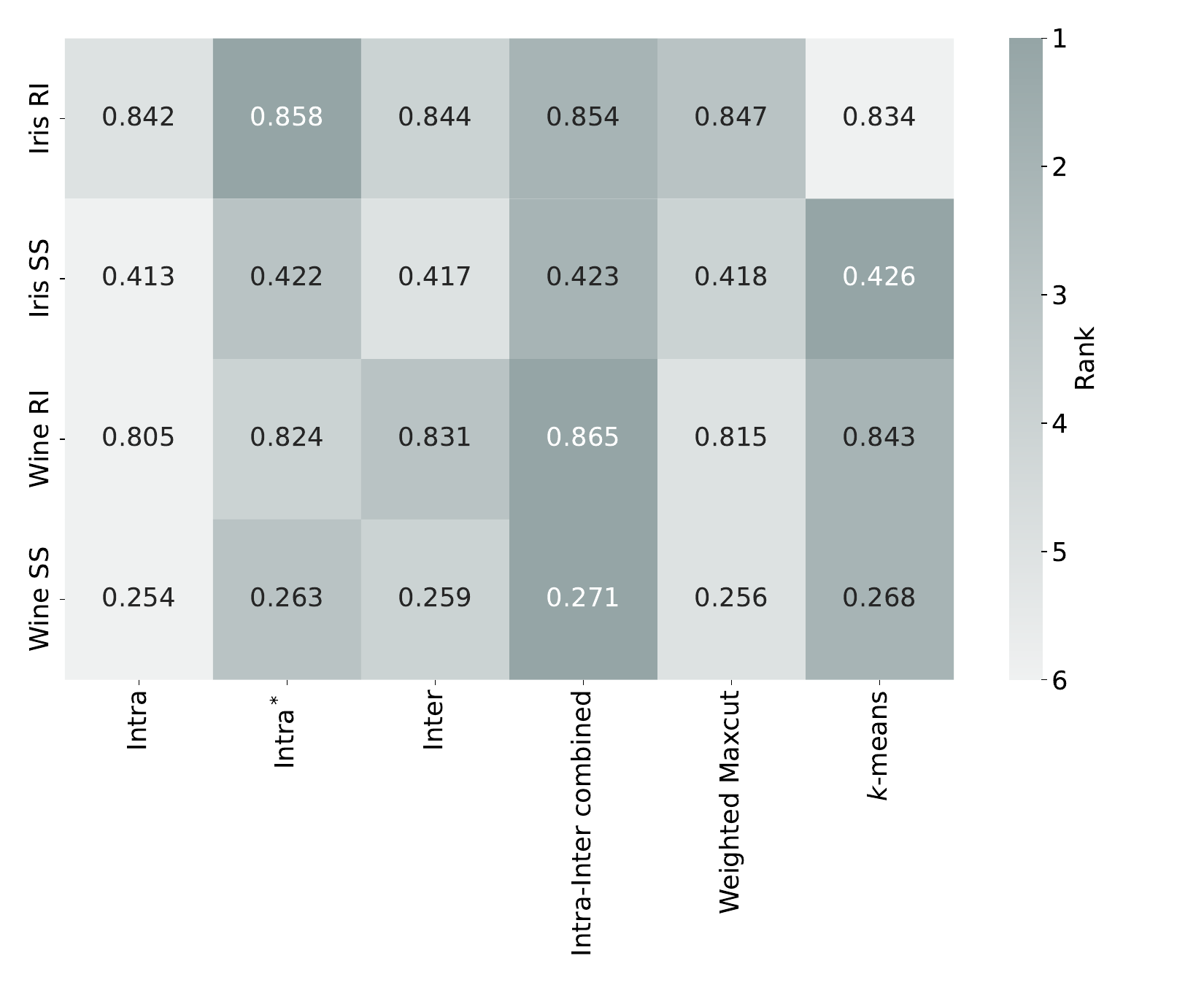}    
    \end{center}
    \caption{The heatmap illustrates exact search results, showing the performance of Hamiltonian methods on the Iris and Wine datasets. The values represent the means of performance metrics (RI : Rand Index, SS : Silhouette Score), with darker shades indicating higher rank (better performance). White text indicates the best results for each evaluation metric.}\label{fig:2}
\end{figure}
To establish a baseline for evaluating the performance of our proposed Hamiltonian methods, we employed a brute-force algorithm to exhaustively search the solution space on small datasets. Although this approach is computationally expensive and infeasible for large datasets, it allows us to find exact solutions and precisely evaluate the performance of our proposed methods. For this reason, we selected the Iris and Wine datasets for our experiments due to their widespread usage as standard benchmarks in clustering and classification tasks, as well as their suitability for exhaustive search given their size. The Iris dataset consists of 150 samples with four features categorized into three classes, while the Wine dataset contains 178 samples with thirteen features also categorized into three classes. To focus on binary clustering, we excluded the Setosa class (50 samples) from the Iris dataset and class 1 (59 samples) from the Wine dataset. For each dataset, we randomly sampled 16 data points and repeated the experiment 150 times. We applied normalization to scale the features of the Iris dataset to a range between 0 and 1. For the Wine dataset, we applied standard scaling to transform the features to have zero mean and unit variance.

Figure~\ref{fig:2} summarizes the performance of different methods on the Iris and Wine dataset. For the Silhouette Score, the $k$-means algorithm achieves the highest score on the Iris dataset. The Intra-Inter combined method and Intra$^*$ method follow closely in second and third place. Notably, the Intra-Inter combined method surpasses the $k$-means algorithm on the Wine dataset. For the Rand Index, one of our Hamiltonian methods outperforms the $k$-means algorithm on both datasets. The Intra$^*$ method achieves the highest Rand Index on the Iris dataset, whereas the Inter method outperforms the Intra$^*$ method on the Wine dataset. By combining Intra and Inter methods, we achieved balanced performance across both datasets. In all cases, at least one of our Hamiltonian methods outperforms the weighted MaxCut, highlighting the benefit of incorporating centroid information into the clustering process.

\begin{figure}[t]
    \begin{center}
    \includegraphics[width=0.5\textwidth]{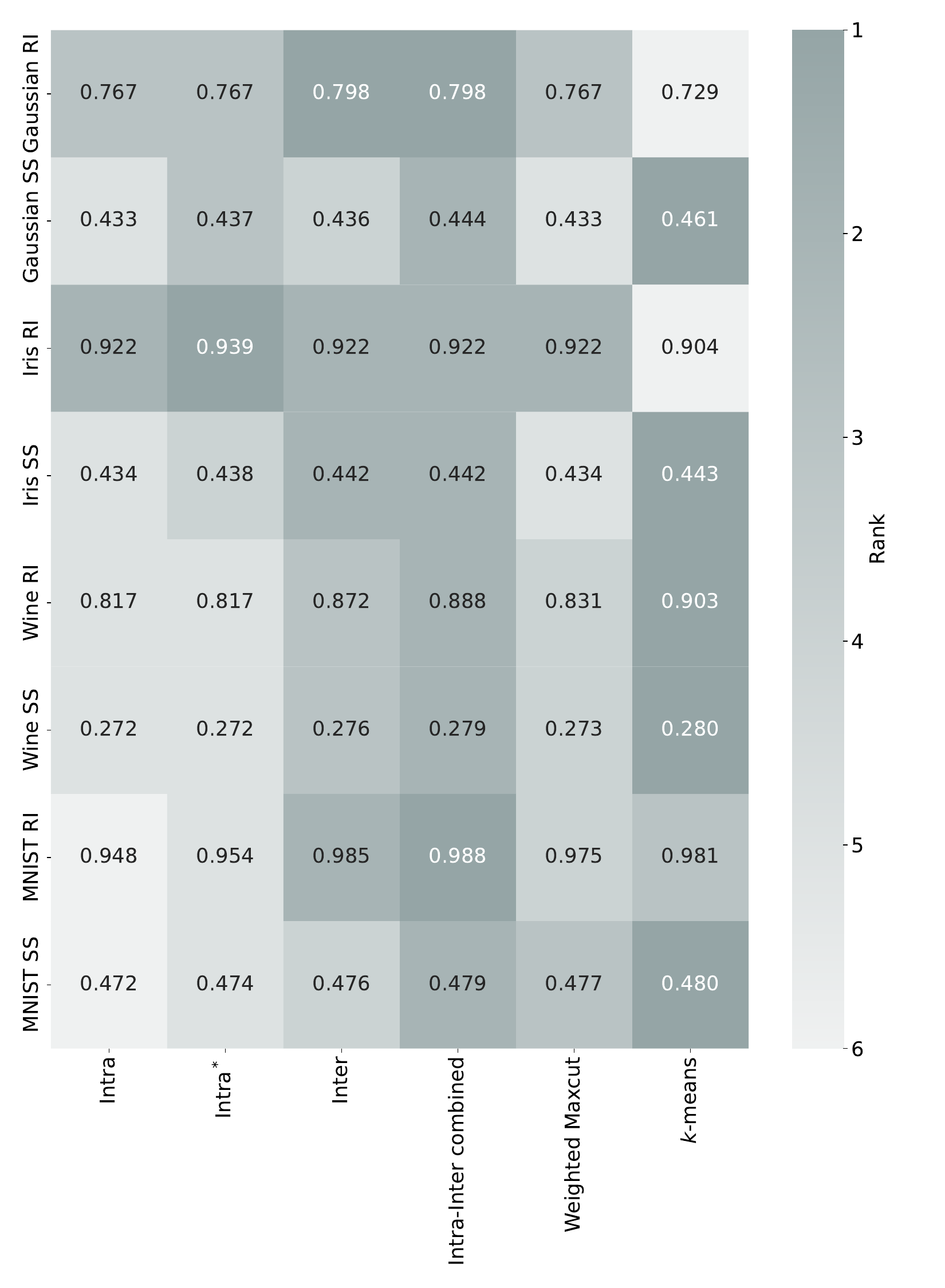}    
    \end{center}
    \caption{The heatmap illustrates simulated annealing results, showing the performance of Hamiltonian methods on the Gaussian synthetic, Iris, Wine and MNIST 0-1 datasets. The values represent the means of performance metrics (RI : Rand Index, SS : Silhouette Score), with darker shades indicating higher rank (better performance). White text highlights the best results for each evaluation metric.}\label{fig:3}
\end{figure}
\subsection{Simulated annealing}
Although the brute-force algorithm guarantees exact solutions, its high computational complexity restricts its application to small datasets. To validate the scalability of our method, we employ the simulated annealing algorithm. This approach enables testing on larger datasets, including not only the Iris and Wine datasets but also Gaussian-distributed synthetic dataset and 0-1 MNIST dataset. The synthetic dataset follows Gaussian distributions with overlapping ranges (see Fig.~\ref{fig:gsoverlap} in Appendix~\ref{sec:Appendix_B}), and the 0-1 MNIST dataset contains handwritten images of digits 0 and 1 in a 28 $\times$ 28 pixel format. We selected 100 samples from the Iris dataset (excluding the Setosa class) and 119 samples from the Wine dataset (excluding class 1). For the Gaussian-distributed synthetic dataset 150 samples were used and 175 samples were chosen from the 0-1 MNIST dataset. All experiments were conducted using an identical annealing schedule, ensuring that each experiment was allocated the same computational time budget.

Figure~\ref{fig:3} presents the performance of different methods across these dataset. For the synthetic dataset, the $k$-means algorithm achieved the highest Silhouette Score compared to other methods. However, both the Inter and Intra-Inter combined methods achieved the highest Rand Index. This indicates that they handle overlapping data more effectively than other methods.
Furthermore, simulations using actual datasets revealed noteworthy results. Although the $k$-means algorithm achieved marginally higher Silhouette Scores, our Hamiltonian methods consistently yielded high Rand Index values while maintaining comparable Silhouette Scores. This enhancement in the Rand Index suggests that our methods not only optimize intracluster cohesion and intercluster separation but also produce cluster assignments that more accurately reflect the true underlying classes. In particular, for the 0-1 MNIST dataset, the Intra-Inter combined method demonstrated excellent performance, indicating that Hamiltonian based clustering can be effectively applied to image recognition tasks.

\subsection{Quantum annealing}

\begin{figure}[t]  
    \centering
    \includegraphics[width=\linewidth]{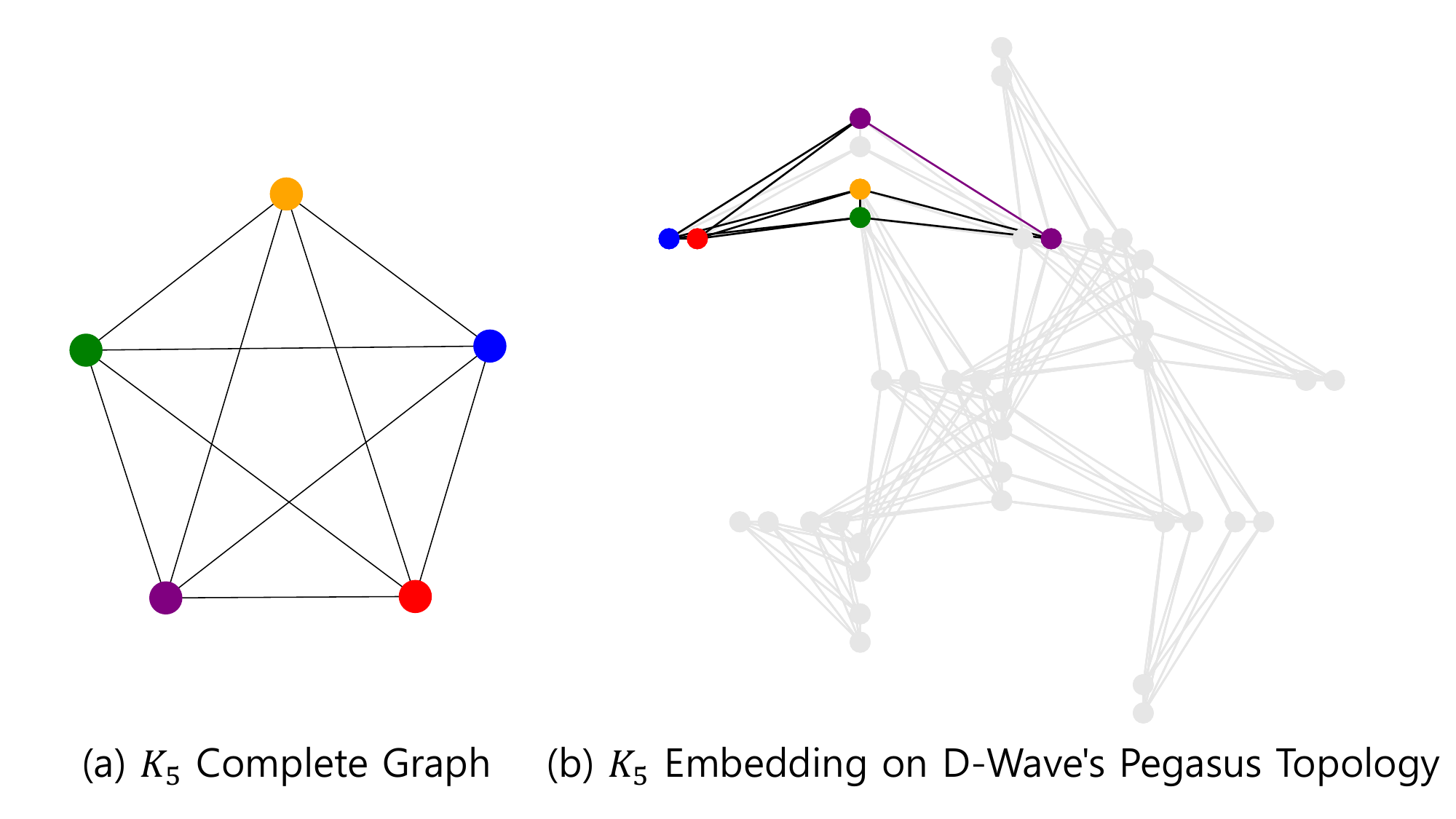}
    \caption{(a) A visualization of a complete graph with 5 vertices, denoted as $K_5$, where each vertex (representing a data point) is connected to every other vertex. (b) The embedding of $K_5$ onto the D-Wave Pegasus topology~\cite{boothby2020nextgenerationtopologydwavequantum}, showing how the fully connected graph is mapped onto the limited qubit connectivity of the quantum hardware. The gray background highlights unused qubits, while colored nodes and edges show the embedded qubits and their connections. The purple node is embedded onto two separate physical qubits, which are linked by a purple edge, representing a chain. This chain ensures that the two qubits act together as a single logical qubit during the quantum annealing process.}
    \label{fig:4}
\end{figure}

To verify that our method can operate on a current quantum device, we performed quantum annealing using the D-Wave Advantage System 6.4, which employs the Pegasus topology (see Fig.~\ref{fig:4} (b)). Our clustering problem inherently involves a fully connected (complete) graph, as depicted in Fig.~\ref{fig:4} (a). This connectivity poses challenges for current quantum devices, which often have limited qubit connectivity. We utilized the clique sampler from the D-Wave Ocean SDK~\cite{boothby2020fast}, which is designed to optimally embed fully connected problems onto the hardware. In graph theory, a clique is a subset of vertices in which every pair of distinct vertices is connected by an edge, forming a complete subgraph. The term clique size refers to the number of vertices in such a fully connected subgraph. Notably, the maximum clique size for the D-Wave Advantage System 6.4 is 175, meaning that it can embed fully connected problems involving up to 175 logical qubits (representing data points). This capability allowed us to process the entire Iris and Wine datasets---each containing fewer than 175 data points---in a single trial. However, the 0-1 MNIST dataset exceeds the limited qubit connectivity of the system, necessitating the random selection of subsets of 175 data points. To ensure statistical robustness, we repeated this sampling process ten times. Our intercluster method demands additional qubits beyond those representing the data points due to the inclusion of higher-order terms (see Eq.~(\ref{ap:inter}) in Appendix~\ref{sec:Appendix_A2}) and hence the slack variables necessary for formulating it as a Binary Quadratic Model (BQM). This extra qubit requirement exceeds the hardware's maximum clique size when handling larger datasets, leading us to exclude the intercluster method from our quantum annealing experiments given the current hardware constraints.

\begin{figure*}[!ht]
  \begin{minipage}[b]{0.5\textwidth}
    \centering
    \includegraphics[width=\linewidth]{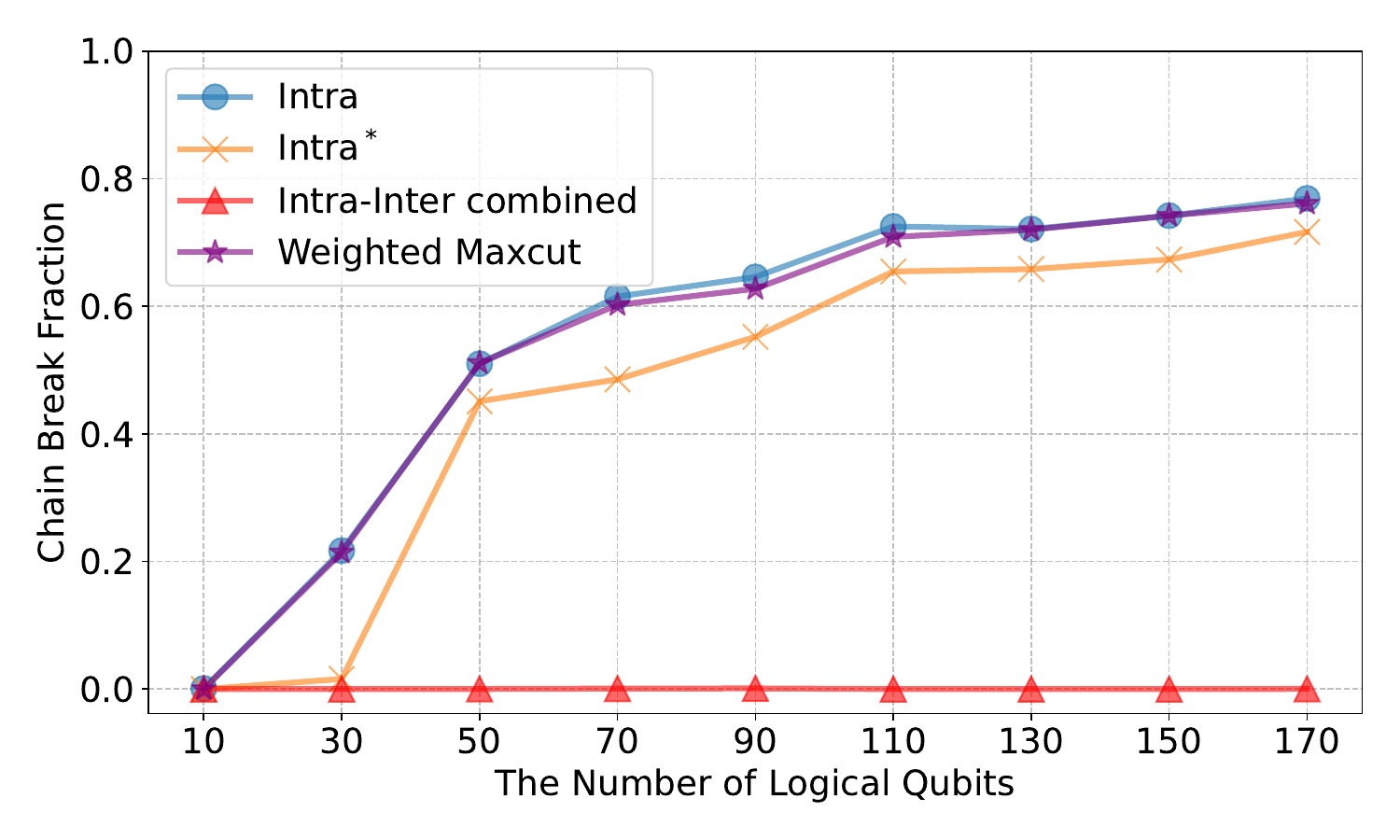}
  \end{minipage}%
  \begin{minipage}[b]{0.5\textwidth}
    \centering
    \includegraphics[width=\linewidth]{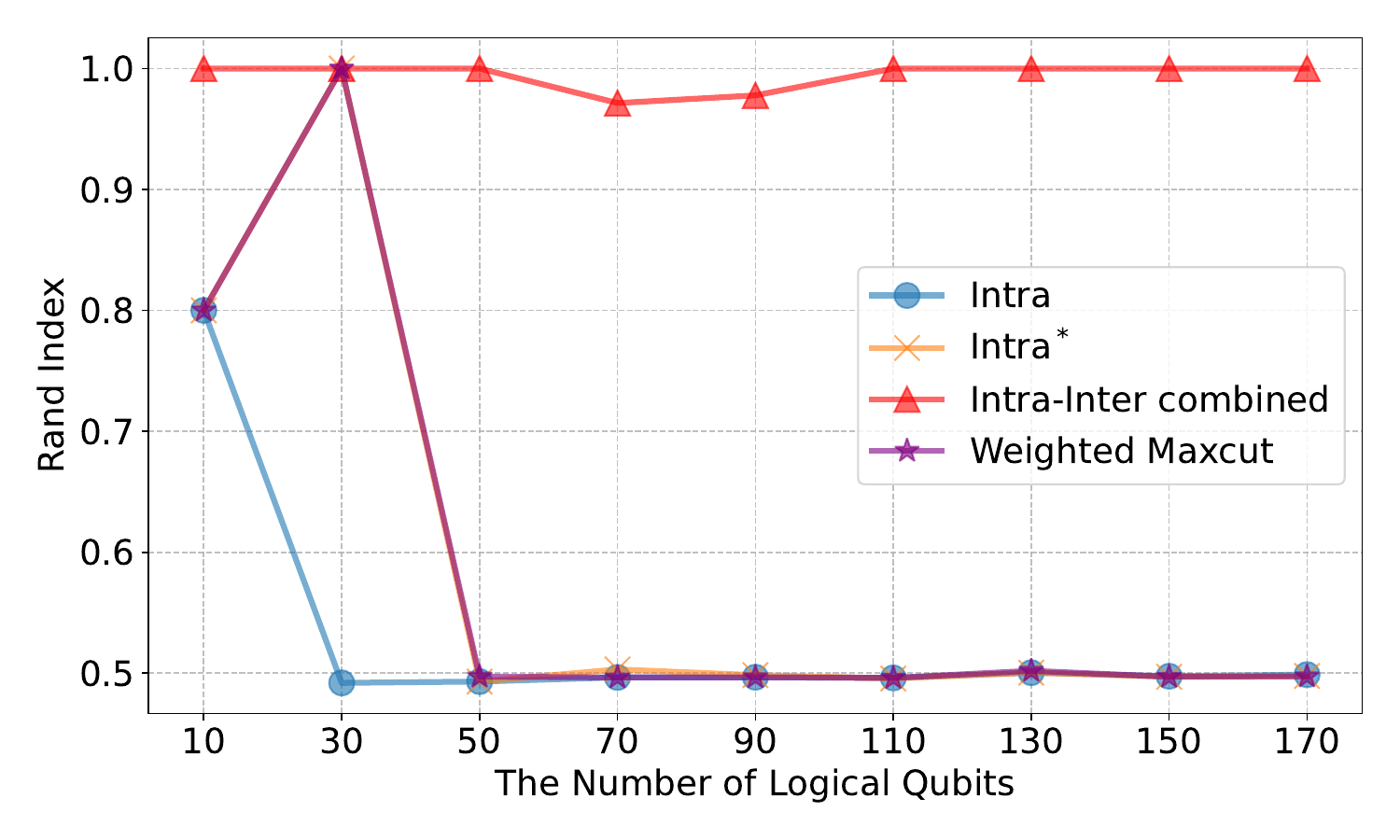}
  \end{minipage}
    \caption{Chain break fraction for four different Hamiltonians using the D-Wave Advantage System 6.4. We conducted 200 samplings on a single quantum machine instruction on a QPU, setting the annealing time to the maximum possible duration of 2000 $\mu$s. The system employed 5612 physical qubits. Using the 0-1 MNIST dataset, we observed chain breaks by averaging the results of 200 samplings. Remarkably, the Intra-Inter combined method did not experience any chain breaks, demonstrating superior stability. This stability significantly influenced the annealing results, resulting in consistently high Rand Index values. When evaluating the Rand Index, we calculated the Rand Index of the minimum energy among the 200 samples, further affirming the robustness and effectiveness of the Intra-Inter combined method.}
    \label{fig:5}
\end{figure*}

\begin{table}[!ht]
\setlength{\tabcolsep}{8pt}
\centering
\begin{tabular}{lcccc}
\toprule
   & \multicolumn{2}{c}{\textbf{Intra-Inter combined}} & \multicolumn{2}{c}{\textbf{$k$-means}} \\ 
\midrule
\textbf{Dataset}   & SS    & RI    & SS    & RI    \\ \midrule
\textbf{Gaussian}  & 0.444 & 0.798 & 0.461 & 0.729 \\
\textbf{Iris}      & 0.442 & 0.922 & 0.443 & 0.904 \\
\textbf{Wine}      & 0.279 & 0.888 & 0.280 & 0.903 \\
\textbf{0-1 MNIST} & 0.479 & 0.989 & 0.480 & 0.981 \\ 
\bottomrule
\end{tabular}
\caption{Comparison of Quantum Annealing results: Silhouette Score and Rand Index between Intra-Inter combined method and $k$-means Algorithm for the Gaussian synthetic, Iris, Wine and 0-1 MNIST datasets.}
\label{tab:QA_comparison}
\end{table}

In this analysis, the Intra-Inter combined method achieved higher Rand Index values while maintaining similar Silhouette Scores compared to the $k$-means algorithm for the Gaussian synthetic, Iris, and 0-1 MNIST datasets, as shown in Table~\ref{tab:QA_comparison}. These results indicate a notable advancement in quantum-enhanced clustering.
In contrast, the Intra, Intra$^*$, and weighted MaxCut methods encountered the logical qubit embedding issue known as chain breaks, leading to random solutions. Figure~\ref{fig:5} illustrates this phenomenon by showing the chain break fractions for four Hamiltonians, highlighting the stability of the Intra-Inter combined method on the 0-1 MNIST dataset. In the process of embedding a problem into the D-Wave Systems, multiple physical qubits are used to represent a single logical qubit. These physical qubits are connected in a chain, as illustrated by the purple edge in Fig.~\ref{fig:4} (b). A chain break occurs when these physical qubits fail to maintain the same state after the annealing process. This misalignment leads to unreliable solutions. Several studies~\cite{hamerly2019experimental,grant2022benchmarking,le2023benchmarking,pelofske2023comparing,gilbert2024quantumannealerschainstrengths} have investigated how chain breaks affect the accuracy of quantum annealing results, highlighting the need for effective embedding strategies and adjustments of chain strength values to minimize such occurrences. Notably, the Intra-Inter combined method performed efficiently on current QPU without the need for additional system parameters tuning or embedding strategies, such as adjusting annealing schedules or chain strengths. 

\subsection{Constrained Clustering}
We performed constrained clustering on the Iris and Wine datasets using Must-Link (ML) and Cannot-Link (CL) constraints, implemented through simulated annealing. Labels of randomly selected data points were revealed according to specified proportions from the entire dataset. Based on these revealed labels, we generated constraints whether pairs of data points should be grouped together (ML) or separated (CL). The proportion of data points with revealed labels ranged from 0\% to 100\% in 10\% increments, resulting in 11 distinct levels. A 0\% ratio reflects a standard clustering scenario without any constraint information, whereas a 100\% ratio indicates that all labels are fully known. For each ratio, we conducted 50 trials using different random samples to calculate average performance metrics. This choice balances statistical robustness and computational efficiency.
\begin{figure*}[!ht]
  \begin{minipage}[b]{0.5\textwidth}
    \centering
    \includegraphics[width=\linewidth]{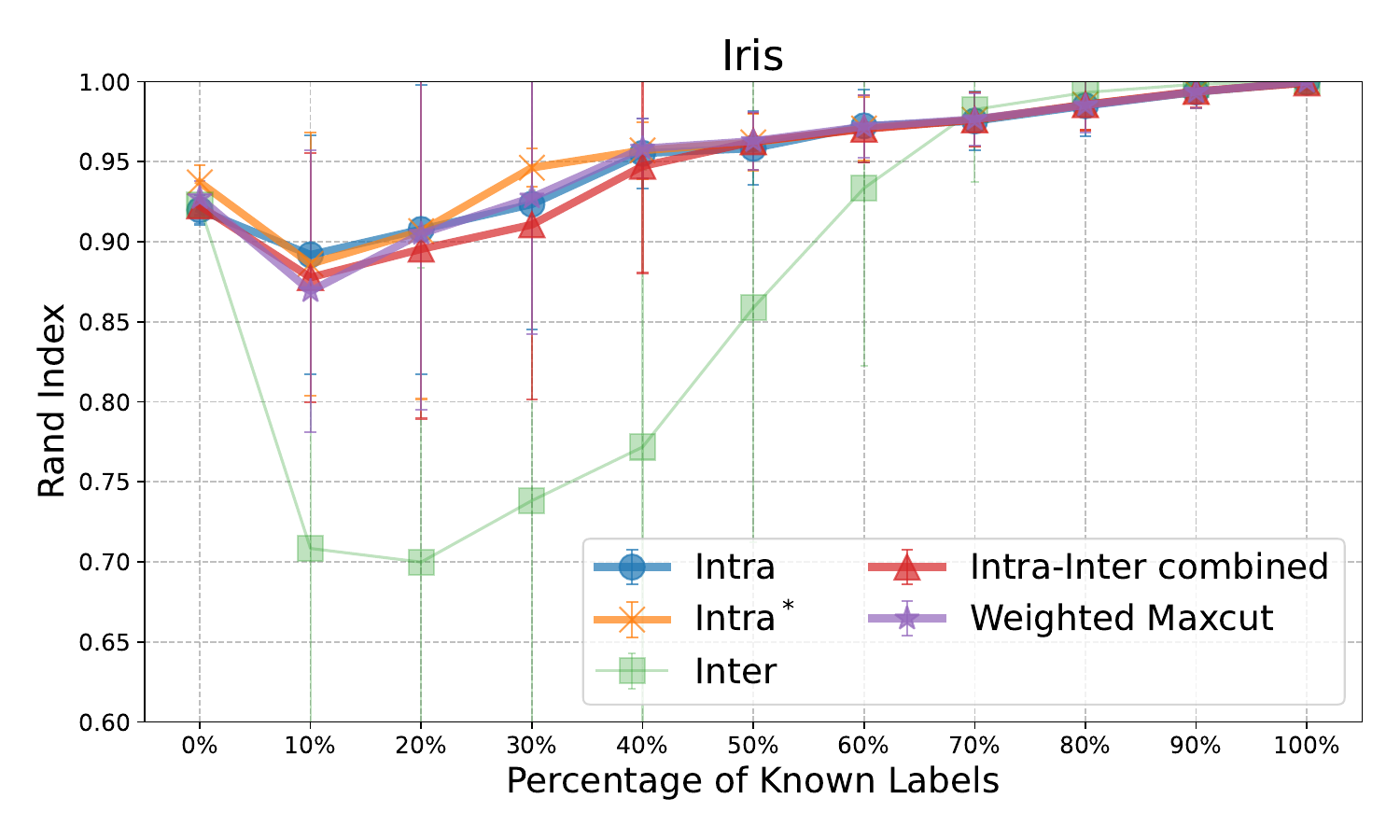}
  \end{minipage}%
  \begin{minipage}[b]{0.5\textwidth}
    \centering
    \includegraphics[width=\linewidth]{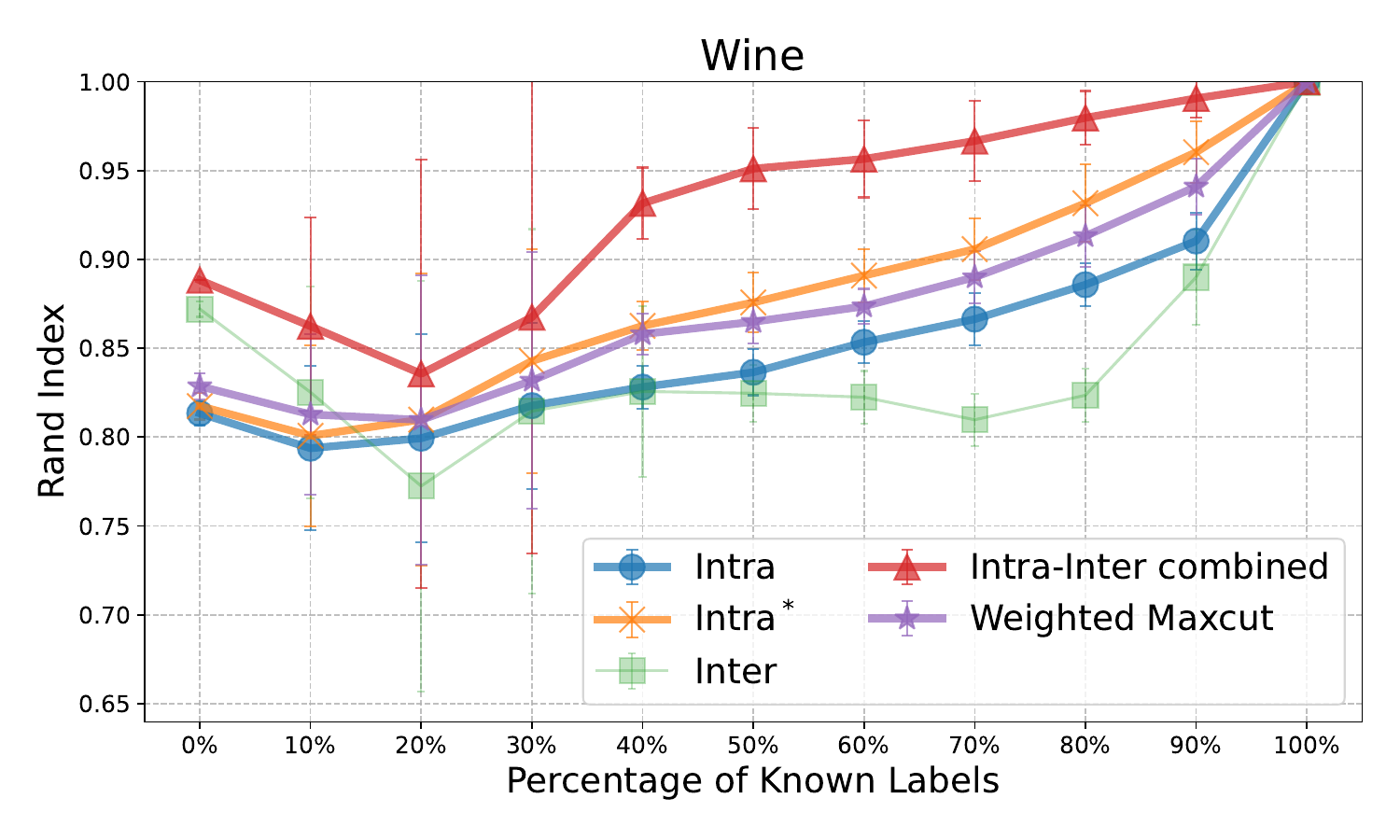}
  \end{minipage}
  \caption{The plots display the Rand Index as a function of the percentage of known labels for the Iris (left) and Wine (right) datasets, under constrained clustering with Must-Link (ML) and Cannot-Link (CL) constraints. Where the proportion of revealed labels ranges from 0\% to 100\% in 10\% increments. The performance improves as more label information is revealed, with the both dataset showing a recovery in performance after an initial decline between 10\% and 30\%. The variability in the Inter method suggests sensitivity to the hyperparameter $\lambda_p$.}
  \label{fig:6}
\end{figure*}

Figure~\ref{fig:6} presents the outcomes of constrained clustering with ML and CL constraints. In both datasets, the Rand Index initially declined when only 10\% to 30\% of label information was provided, but progressively improved as more information available. In the case of Wine dataset, the Intra-Inter combined method quickly converged to the true labels. In contrast, the Inter method showed no consistent trend and exhibited considerable variability. This suggests that the solution values are sensitive to the hyperparameter $\lambda_p$, which can influence the prioritization of constraints or clustering methods.

Subsequently, we implemented cardinality constraints to assign a specific number of data points to each cluster on the Iris and Wine datasets using simulated annealing. For each experiment, we selected a total of 50 data points from the Iris dataset and 48 data points from the Wine dataset, drawn from labels 1 and 2 according to predetermined ratios. The proportions of data points from label 1 and label 2 varied as follows: (10\%, 90\%), (20\%, 80\%), (30\%, 70\%), (40\%, 60\%), (50\%, 50\%), (60\%, 40\%), (70\%, 30\%), (80\%, 20\%), and (90\%, 10\%). This means we started with 10\% data points from label 1 and 90\% from label 2, gradually adjusting the proportions until we reached 90\% from label 1 and 10\% from label 2.
\begin{figure*}[!ht]
    \includegraphics[width=\linewidth]{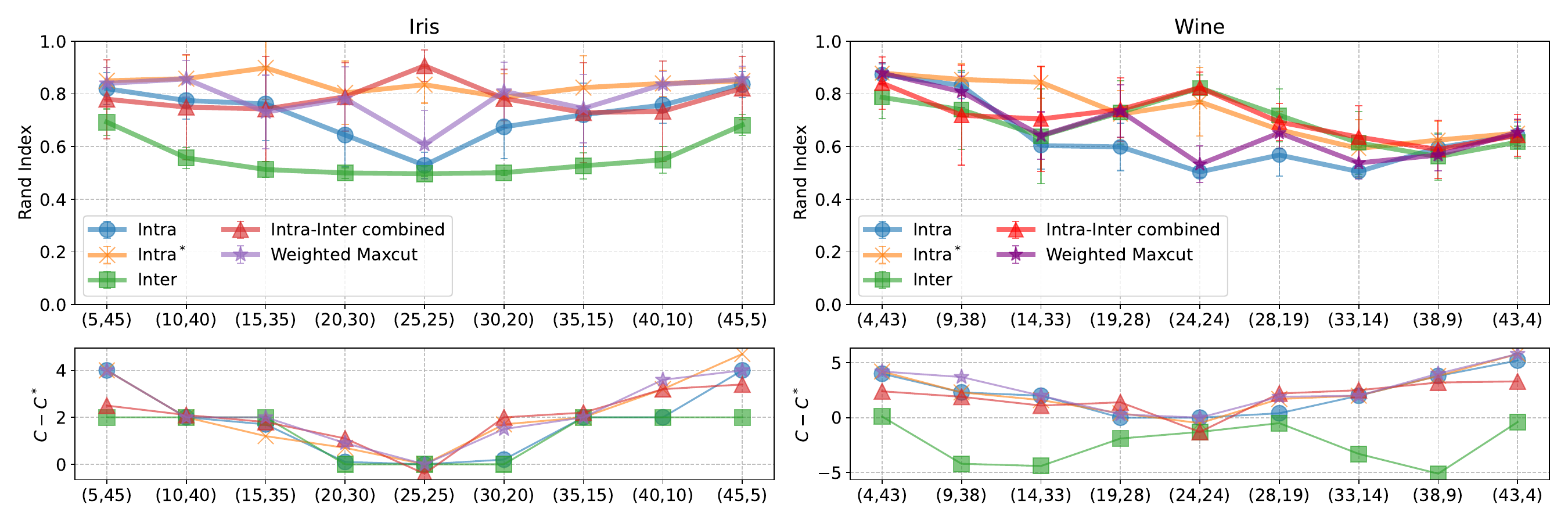}
    \caption{The top plots show the Rand Index across various methods with cardinality constraints applied to the Iris (left) and Wine (right) datasets. The bottom plots depict the difference between the given cardinality ($C$) and the experimental results ($C^*$). The horizontal axis represents the number of data points from label 1 to label 2, ranging from (10\%, 90\%) to (90\%, 10\%). Notice that the Inter method underperformed on the Iris dataset. In contrast, the Intra$^*$ and Intra-Inter combined methods maintained a high Rand Index while closely achieving the desired cardinality.}
    \label{fig:7}
\end{figure*}

Figure~\ref{fig:7} illustrates the performance of our methods with cardinality constraints. The Intra$^*$ and Intra-Inter combined methods achieved high Rand Index values across different label ratios for both datasets, closely matching the desired cardinality values. On the Iris dataset, the Intra, weighted MaxCut and Inter methods exhibited less satisfactory performance with balanced data points (50\% from each label). On the Wine dataset, the Intra and weighted MaxCut methods also underperformed with balanced data points. Nevertheless, the results of the Intra$^*$ and Intra-Inter combined methods indicate that it is possible to perform clustering on imbalanced data and adjust cluster sizes according to user specifications, effectively addressing real-world clustering problems.

\section{Conclusions and Discussion}
In this work, we formulated the clustering problem as finding the ground state of a Hamiltonian and developed methods to integrate centroid information directly into the objective function. We defined a distance function $l(\mu,z,s)$ that encompasses intracluster distance, intercluster distance, and a combination of both. By incorporating the number of data points in each cluster as a variable within the objective function, we eliminated the need for fixed cluster size assumptions. We also extended our method to constrained clustering, enabling domain experts to embed prior knowledge into the clustering process. Our experimental results demonstrated that at least one of our proposed Hamiltonians outperforms the weighted MaxCut across multiple datasets, including both synthetic and real-world examples. This underscores the importance of incorporating centroid information in clustering algorithms. Notably, the Intra-Inter combined method exhibited balanced performance across most datasets.

The significance of our research lies in developing a flexible and unified clustering strategy capable of addressing complex clustering challenges. By enabling the integration of various clustering objectives, our approach effectively manages data points clustered around their mean and handles overlapping clusters, as evidenced by our experimental results. Application to real datasets further highlights the practical utility of the Hamiltonian formulation. A key benefit of our method is its compatibility with quantum simulation techniques. In particular, our quantum annealing experiments on the D-Wave Systems showed that the Intra-Inter combined method operates effectively on current quantum device, demonstrating its applicability to real-world problems. 

Potential directions for future research include expanding the Hamiltonian-based clustering framework to develop data-driven, automated methods for determining the optimal number of clusters. The intracluster distance formulation in our Hamiltonian approach can be leveraged to enforce density constraints. Additionally, refining the Intra-Inter combined method by applying dynamic weights to the linear combination of distances could allow adaptation to context-specific requirements. Extending this Hamiltonian formulation to address more complex clustering scenarios, such as time series and high-dimensional datasets, also presents a promising direction. Such advancements have the potential to enable new applications in fields like finance, drug discovery, and social network analysis.

\section*{Acknowledgments}
This work was supported by Korea Research Institute for Defense Technology Planning and Advancement---grant funded by Defense Acquisition Program Administration (DAPA) (KRIT-CT-23-031). This work was also supported by Institute of Information \& communications Technology Planning \& evaluation (IITP) grant funded by the Korea government (No. 2019-0-00003, Research and Development of Core technologies for Programming, Running, Implementing and Validating of Fault-Tolerant Quantum Computing System), the Yonsei University Research Fund of 2024 (2024-22-0147), the National Research Foundation of Korea (2023M3K5A1094813), and the KIST Institutional Program (2E32941-24-008). Access to the D-Wave system was supported by the `Quantum Information Science R\&D Ecosystem Creation' through the National Research Foundation of Korea (NRF) funded by the Korean government (Ministry of Science and ICT) (2020M3H3A1110365).

\clearpage
\appendix
\onecolumngrid
\section{Objective functions revisited}
\label{sec:Appendix_A}

In this section, we dissect the mathematical formulations introduced in the main text. We begin by expanding the terms $l(\mu_{\pm},z,\pm 1)$, which serve as the core components for constructing the various objective functions, as follows:
\begin{equation}\label{1st_term_raw}
        l(\mu_+,z,+1) = \sum_{j=1}^d \left[\sum_{i=1}^{N} {x_i}_j^2 \frac{1+z_i}{2} - \frac{2}{N_+} \left(\sum_{i=1}^N {x_i}_j \frac{1+z_i}{2}\right)^2 + \frac{N_+}{N_+^2}\left(\sum_{i=1}^N {x_i}_j \frac{1+z_i}{2}\right)^2 \right].
    \end{equation}

\begin{equation}\label{2nd_term_raw}
    l(\mu_-,z,-1)
    = \sum_{j=1}^d \left[\sum_{i=1}^{N} {x_i}_j^2 \frac{1-z_i}{2} - \frac{2}{N_-} \left(\sum_{i=1}^N {x_i}_j \frac{1-z_i}{2}\right)^2 + \frac{N_-}{N_-^2}\left(\sum_{i=1}^N {x_i}_j \frac{1-z_i}{2}\right)^2 \right].
\end{equation}

\begin{equation}\label{3rd_term_raw}
    - l(\mu_-,z,+1)
    =\sum_{j=1}^d\left[-\sum_{i=1}^{N}{{x_i}_j^2 \frac{1+z_i}{2}} + \frac{2}{N_-}{\sum_{i=1}^{N}{x_i}_j \frac{1+z_i}{2}}{\sum_{i=1}^{N}{x_i}_j \frac{1-z_i}{2}} - \frac{N_+}{N_-^2}\left(\sum_{i=1}^{N}{x_i}_j \frac{1-z_i}{2}\right)^2\right].
\end{equation}

\begin{equation}\label{4th_term_raw}
    - l(\mu_+,z,-1)
    =\sum_{j=1}^d\left[-\sum_{i=1}^{N}{{x_i}_j^2 \frac{1-z_i}{2}} + \frac{2}{N_+}{\sum_{i=1}^{N}{x_i}_j \frac{1+z_i}{2}}{\sum_{i=1}^{N}{x_i}_j \frac{1-z_i}{2}} - \frac{N_-}{N_+^2}\left(\sum_{i=1}^{N}{x_i}_j \frac{1+z_i}{2}\right)^2\right].
\end{equation}
Here, $x_{ij}$ represents the $j$th entry (or feature) of the $i$th data vector $x_i$. These expressions explicitly show that the terms $N_+$ and $N_-$ appear in the denominators with powers of either 1 or 2, indicating the necessary factors to multiply for transforming these equations into the Hamiltonian form.
In the following subsections, we reorganize and simplify the clustering objective functions introduced in Sec.~\ref{sec:intra} through Sec.~\ref{sec:intrainter} of the main manuscript and present the corresponding Hamiltonians to provide additional insights into the optimization framework.

\subsection{Intracluster distance}
\label{sec:Appendix_A1}
To capture the intracluster distance, we combine the following equations Eqs.~\eqref{1st_term_raw} and ~\eqref{2nd_term_raw}:
\begin{equation}\label{ap:denominator}
    \sum_{j=1}^d \left[\sum_{i=1}^{N} {{x_i}_j}^2 - \frac{N_+N_-^2}{N_+^2N_-^2}\left(\sum_i^N {x_i}_j\frac{1+z_i}{2}\right)^2 - \frac{N_+^2N_-}{N_+^2N_-^2}\left(\sum_i^N {x_i}_j\frac{1-z_i}{2}\right)^2 \right].
\end{equation}
Multiplying by $N_+^2N_-^2$ to eliminate the denominator yields:
\begin{equation}\label{ap:intra_raw}
    \sum_{j=1}^d \left[N_{+}^2N_{-}^2\sum_{i=1}^{N} {{x_i}_j}^2 - N_{+}N_{-}^2\left(\sum_i^N {{x_i}_j}\frac{1+z_i}{2}\right)^2  - N_{+}^2N_{-}\left(\sum_i^N {{x_i}_j}\frac{1-z_i}{2}\right)^2 \right].
\end{equation}
However, we observe that Eq.~\eqref{ap:intra_raw} collapses to zero when all data points are assigned to a single cluster (i.e. $N_+=0$ or $N_-=0$), resulting in a trivial solution where the objective function reaches its minimum, making this outcome undesirable.
To overcome this limitation, we multiply Eq.~\eqref{1st_term_raw} by $N_+^2$ and Eq.~\eqref{2nd_term_raw} by $N_-^2$:
\begin{equation}
    \sum_{j=1}^d \left[N_+^2 \sum_{i=1}^N {x_i}_j^2 \frac{1+z_i}{2} - N_+ \left(\sum_{i=1}^N {x_i}_j \frac{1+z_i}{2}\right)^2\right].
\end{equation}
\begin{equation}
    \sum_{j=1}^d \left[N_-^2 \sum_{i=1}^N {x_i}_j^2 \frac{1-z_i}{2} - N_- \left(\sum_{i=1}^N {x_i}_j \frac{1-z_i}{2}\right)^2\right].
\end{equation}
By combining these two equations and simplifying, we obtain the following result:
\begin{equation}
    \sum_{k=1}^d \left[\frac{N^2}{4}\sum_{i=1}^N {x_i}_k^2 - \frac{N}{4}\sum_{i=1}^N {x_i}_k^2 - \frac{N}{4}\left(\sum_{i=1}^N {x_i}_k \right)^2 + \frac{N}{4}\left(N_{+}-N_{-}\right)\sum_{i=1}^N{x_i}_k^2z_i + \frac{N_{+}-N_{-}}{4}\sum_{i=1}^N{x_i}_k^2\sum_{i=1}^Nz_i \right.\nonumber
\end{equation}
\begin{equation}
    \left.- \frac{N_{+}-N_{-}}{2}\sum_{i=1}^N{x_i}_k\sum_{i=1}^N{x_i}_k z_i + \frac{N}{4}\sum_{i=1}^N z_i \sum_{i=1}^N{x_i}_k^2z_i - \frac{N}{2}\sum_{i<j}^N {x_i}_k {x_j}_k z_iz_j \right].
\end{equation}
The objective function becomes:
\begin{equation}
    \sum_{k=1}^d \left[\frac{N}{2}\sum_{i<j}^N {x_i}_k^2 z_iz_j + \frac{N}{2}\sum_{i<j}^N {x_j}_k^2z_iz_j + \frac{1}{2}\sum_i^N{x_i}_k^2\sum_{i<j}^Nz_iz_j \right. \nonumber
\end{equation}
\begin{equation}
    \left.-\frac{1}{2}\sum_i^N{x_i}_k\sum_{i<j}^N\left({x_i}_k+{x_j}_k\right)z_iz_j -\frac{N}{2}\sum_{i<j}^N{x_i}_k{x_j}_kz_iz_j \right]
\end{equation}
\begin{equation}\label{ap:intras}
    =\sum_{k=1}^d \sum_{i<j}^N \left[\frac{N}{2}{x_i}_k^2 + \frac{N}{2}{x_j}_k^2 + \frac{1}{2}\sum_{i=1}^N{x_i}_k^2-\frac{1}{2}\left(\sum_{i=1}^N{x_i}_k\right)\left({x_i}_k+{x_j}_k\right)-\frac{N}{2}{x_i}_k{x_j}_k\right]z_iz_j.
\end{equation}
Equation~\eqref{ap:intras} corresponds to Eq.~\eqref{eq:intra2} in the main manuscript. The Hamiltonian $H$ for the intracluster optimization problem can then be derived using a similar procedure as outlined in Eqs.~(\ref{eq:qubo}) to ~(\ref{eq:maxcut}), as follows:
\begin{equation}
    H=\sum_{k=1}^d \sum_{i<j}^N \left[\frac{N}{2}{x_i}_k^2 + \frac{N}{2}{x_j}_k^2 + \frac{1}{2}\sum_i^N{x_i}_k^2-\frac{1}{2}\left(\sum_i^N{x_i}_k\right)\left({x_i}_k+{x_j}_k\right)-\frac{N}{2}{x_i}_k{x_j}_k\right]Z_iZ_j.
\end{equation}
Alternatively, in Eq.~\eqref{ap:denominator}, after canceling terms, we observe that the denominator has a power of 1. To analyze how different powers of $N_{\pm}$ influence the clustering results, we multiply Eq.~\eqref{1st_term_raw} by $N_+$ and Eq.~\eqref{2nd_term_raw} by $N_-$:
\begin{equation}\label{eq42}
    \sum_{j=1}^d \left[N_{+}\sum_{i=1}^{N} {x_i}_j^2 \frac{1+z_i}{2} - 2 \left(\sum_{i=1}^N {x_i}_j \frac{1+z_i}{2}\right)^2 + \left(\sum_{i=1}^N {x_i}_j \frac{1+z_i}{2}\right)^2 \right].
\end{equation}
\begin{equation}\label{eq43}
        \sum_{j=1}^d \left[N_{-}\sum_{i=1}^{N} {x_i}_j^2 \frac{1-z_i}{2} - 2\left(\sum_{i=1}^N {x_i}_j \frac{1-z_i}{2}\right)^2 + \left(\sum_{i=1}^N {x_i}_j \frac{1-z_i}{2}\right)^2 \right].
\end{equation}
By combining these two equations and simplifying, we obtain the following result:
\begin{equation}
    \sum_{k=1}^d\left[\frac{N}{2}\sum_{i}^N {x_i}_k^2 + \frac{1}{2} \sum_{i=1}^{N} z_i \sum_{i=1}^N {x_i}_k^2 z_i - \frac{1}{2}\left(\sum_{i=1}^N {x_i}_k\right)^2 -\frac{1}{2}\sum_{i}^N {x_i}_k^2 - \sum_{i<j}^N {x_i}_k{x_j}_k  z_i z_j\right].
\end{equation}
The objective function becomes:
\begin{equation}
    \sum_{k=1}^d \frac{1}{2}\sum_{i<j}^N {x_i}_k^2 z_iz_j + \frac{1}{2}\sum_{i<j}^N {x_j}_k^2z_iz_j - \sum_{i<j}^N {x_i}_k{x_j}_k z_i z_j
    \label{ap:intra}
    = \sum_{k=1}^d \sum_{i<j}^N \frac{1}{2} ({x_i}_k-{x_j}_k)^2 z_i z_j.
\end{equation}
Equation~\eqref{ap:intra} corresponds to combining $l(\mu_+,z,+1)$ and $l(\mu_-,z,-1)$ with the multiplicative factor $N_+$ and $N_-$. The Hamiltonian $H$ for this optimization problem can be derived using a similar procedure as outlined in Eqs.~(\ref{eq:qubo}) to ~(\ref{eq:maxcut}), as follows:
\begin{equation}
    H = \sum_{k=1}^d \sum_{i<j}^N \frac{1}{2} ({x_i}_k-{x_j}_k)^2 Z_i Z_j.
\end{equation}

\subsection{Intercluster distance}
\label{sec:Appendix_A2}
To capture the intercluster distance, we combine Eq.~\eqref{3rd_term_raw} and Eq.~\eqref{4th_term_raw}. After simplifying, we get:
\begin{equation}
    \sum_{j=1}^d \left[-\sum_{i=1}^N {x_i}_j^2 + \frac{2N}{N_+ N_-} \sum_{i=1}^N {x_i}_j \frac{1+z_i}{2} \sum_{i=1}^N {x_i}_j \frac{1-z_i}{2} - \frac{N_+}{N_-^2} \left(\sum_{i=1}^N \frac{1-z_i}{2}\right)^2 - \frac{N_-}{N_+^2}\left(\sum_{i=1}^N {x_i}_j \frac{1+z_i}{2}\right)^2\right].
\end{equation}
Multiplying by $N_+^2N_-^2$ to eliminate the denominators, we obtain the following objective function:
\begin{equation}
    \sum_{j=1}^d \left[-{N_{+}^2N_{-}^2}\sum_{i=1}^{N}{{x_i}_k^2} + 2NN_{+}N_{-}{\sum_{i=1}^{N}{x_i}_k \frac{1+z_i}{2}}{\sum_{i}^{N}{x_i}_k \frac{1-z_i}{2}} \right.\nonumber\\
\end{equation}
\begin{equation}
    \left.- {N_{+}^3}\left(\sum_{i=1}^{N}{x_i}_k \frac{1-z_i}{2}\right)^2 - {N_{-}^3}\left(\sum_{i}^{N}{x_i}_k \frac{1+z_i}{2}\right)^2\right].
\end{equation}
To facilitate the representation of interactions between data points, we define strict upper triangular matrix $U_s$, $X_s^{(k)}$, and vector $\mathbf{1}$, where $U_s \in \mathbb{R}^{N\times N}$, $X_s^{(k)} \in \mathbb{R}^{N \times N}$, and $\mathbf{1}\in \mathbb{R}^N$:
\begin{equation}
            U_s = 
    \begin{bmatrix}
    0 & 1 & \cdots & 1 \\
    0 & 0 & \cdots & 1 \\
    \vdots  & \vdots  & \ddots & \vdots  \\
    0 & 0 & \cdots & 0
    \end{bmatrix}, \quad
    {X_s}^{(k)} = 
    \begin{bmatrix}
    0 & {x_1}_k{x_2}_k & \cdots & {x_1}_k{x_N}_k \\
    0 & 0 & \cdots & {x_2}_k{x_N}_k \\
    \vdots  & \vdots  & \ddots & \vdots  \\
    0 & 0 & \cdots & 0
    \end{bmatrix}, \quad
    \mathbf{1}=
    \begin{bmatrix}
    1 & 1 & \cdots & 1
    \end{bmatrix}^T.
\end{equation}
Using these matrices and vectors, the objective function can be expressed as:
\begin{equation}
    \sum_{k=1}^d \left[\left(\frac{N^2}{4}-\frac{3}{8}N\right){x_k}^Tx_k - \frac{5}{8}N(\mathbf{1}^T{x_k})^2\right]z^TU_sz - \frac{1}{4}{x_k}^T{x_k}(z^TU_sz)^2-(\frac{3}{8}N^3+\frac{N^2}{8})z^TX_s^{(k)}z\nonumber
\end{equation}
\begin{equation}\label{ap:inter}
    + \left(\frac{3}{8}N^2+\frac{N}{8}\right)z^T({x_k}{x_k}^T\mathbf{1}\mathbf{1}^T)z -\frac{N}{4}z^TU_szz^TX_s^{(k)}z + \frac{1}{4}z^T({x_k}{x_k}^T\mathbf{1}\mathbf{1}^T)zz^TU_sz.
\end{equation}
Equation~\eqref{ap:inter} corresponds to combining $-l(\mu_-,z,+1)$ and $-l(\mu_+,z,-1)$ with the multiplicative factor $N_{+}^2 N_{-}^2$, as shown in Eq.~\eqref{eq:inter} in the main manuscript.
The objective function contains higher-order terms (e.g. $z^TU_szz^TX_s^{(k)}z$), so slack variables must be used for the D-Wave Systems. Therefore, the intercluster method was not implemented on the quantum annealing experiments, and the results were verified using simulated annealing and brute-force algorithms. This is due to the hardware limitations of existing quantum annealers, which are more suited to quadratic terms.

\subsection{Combining intra and intercluster distances}
\label{sec:Appendix_A3}
To capture the intra and intercluster distances, we combine the following equations Eqs.~\eqref{1st_term_raw} + ~\eqref{2nd_term_raw} + ~\eqref{3rd_term_raw} + ~\eqref{4th_term_raw}:
\begin{equation}
    \sum_{j=1}^d -\frac{1}{N_+} \left(\sum_{i=1}^N {x_i}_j \frac{1+z_i}{2}\right)^2 - \frac{1}{N_-} \left(\sum_{i=1}^N {x_i}_j \frac{1-z_i}{2}\right)^2 +\frac{2N}{N_+N_-}\left(\sum_{i=1}^N {x_i}_j \frac{1+z_i}{2}\right)\left(\sum_{i=1}^N {x_i}_j \frac{1-z_i}{2}\right) \nonumber
\end{equation}
\begin{equation}
    - \frac{N_+}{N_{-}^2} \left(\sum_{i=1}^N {x_i}_j \frac{1-z_i}{2}\right)^2 -\frac{N_-}{N_{+}^2}\left(\sum_{i=1}^N {x_i}_j \frac{1+z_i}{2}\right)^2.
\end{equation}
\begin{equation}\label{ap:intrainter_dissect}
    =\sum_{j=1}^d -\frac{N}{N_+^2 N_-^2}\left[N_{-}^2 \left(\sum_{i=1}^N {x_i}_j \frac{1+z_i}{2}\right)^2 - 2N_{+1}N_{-1}\left(\sum_{i=1}^N {x_i}_j \frac{1+z_i}{2}\right)\left(\sum_{i=1}^N {x_i}_j \frac{1-z_i}{2}\right) 
    +N_{+}^2 \left(\sum_{i=1}^N {x_i}_j \frac{1-z_i}{2}\right)^2 \right].
\end{equation}
By simplifying Eq.~\ref{ap:intrainter_dissect}, the objective function can be expressed as:
\begin{equation}
    \sum_{j=1}^d -\frac{N}{N_+^2 N_-^2} \left(N_- \sum_{i=1}^N {x_i}_j \frac{1+z_i}{2} - N_+ \sum_{i=1}^N {x_i}_j \frac{1-z_i}{2}\right)^2.
\end{equation}
Multiplying by $N_+^2 N_-^2$ to eliminate the denominators, we get:
\begin{equation}
    \sum_{j=1}^d -N\left(N_{-}\sum_{i=1}^N{{x_i}_j}\frac{1+z_i}{2}-N_{+}\sum_{i=1}^{N}{{x_i}_j}\frac{1-z_i}{2}\right)^2.
\end{equation}
Expanding and simplifying, the objective function can be expressed as:
\begin{equation}\label{ap:intrainter}
    \sum_{k=1}^d\left[-\frac{1}{2}\left(\sum_i^N{x_i}_k\right)^2\sum_{i<j}^N{z_iz_j}+\frac{N}{2}\sum_i^N{x_i}_k\sum_{i<j}^N{{x_i}_kz_iz_j}+\frac{N}{2}\sum_i^N{x_i}_k\sum_{i<j}^N{{x_j}_kz_iz_j}-\frac{N^2}{2}\sum_{i<j}^N{{x_i}_k{x_j}_kz_iz_j}\right].
\end{equation}
Equation~\eqref{ap:intrainter} corresponds to linearly combining intracluster distance and intercluster distance with the multiplicative factor $N_{+}^2 N_{-}^2$, as shown in Eq.~\eqref{eq:intra and inter}. The Hamiltonian $H$ for the intra and intercluster optimization problem can then be derived using a similar procedure as outlined in Eqs.~(\ref{eq:qubo}) to ~(\ref{eq:maxcut}), as follows:
\begin{equation}
    H = \sum_{k=1}^d \sum_{i<j}^N \left[ -\frac{1}{2}\left(\sum_{i}^N{x_i}_k\right)^2 + \frac{N}{2}\left(\sum_{i=1}^N{x_i}_k\right)({{x_i}_k + {x_j}_k}) - \frac{N^2}{2}{{x_i}_k}{{x_j}_k}\right]Z_iZ_j.
\end{equation}

\section{Experiments Results}
\label{sec:Appendix_B}
This section contains experimental outcomes with additional metrics. We employ three metrics to evaluate clustering performance: Dist Centroid, which measures the separation betweeen cluster centroids; Intra, the sum of intracluster distances; and Inter, the sum of intercluster distances. Larger values for Dist Centroid and Inter, and lower values for Intra, indicate better clustering. In the tables, the best-performing method is highlighted in bold, and the top result among the Hamiltonian methods is underlined.

\begin{table*}[!ht]
    \centering
    \fontsize{9}{11}\selectfont
    \begin{tabular}{lcccccc}
    \toprule
      & Intra & Intra$^*$ & Inter & Intra-Inter combined & Weighted MaxCut & $k$-means \\
    \midrule
    \multicolumn{7}{l}{\textbf{Iris Dataset}} \\
    RI & 0.842 $\pm$ 0.123 & \underline{\textbf{0.858 $\pm$ 0.123}} & 0.844 $\pm$ 0.130 & 0.854 $\pm$ 0.130 & 0.847 $\pm$ 0.127 & 0.834 $\pm$ 0.144 \\
    SS & 0.413 $\pm$ 0.074 & 0.422 $\pm$ 0.074 & 0.417 $\pm$ 0.075 & \underline{0.423 $\pm$ 0.074} & 0.418 $\pm$ 0.073 & \textbf{0.426 $\pm$ 0.078} \\
    Dist Centroid & 0.094 $\pm$ 0.013 & 0.094 $\pm$ 0.012 & 0.094 $\pm$ 0.013 & \underline{0.095 $\pm$ 0.013} & 0.094 $\pm$ 0.013 & \textbf{0.096 $\pm$ 0.013} \\
    Intra & 0.645 $\pm$ 0.071 & 0.640 $\pm$ 0.071 & 0.642 $\pm$ 0.070 & \underline{\textbf{0.639 $\pm$ 0.070}} & 0.642 $\pm$ 0.070 & 0.643 $\pm$ 0.077 \\
    Inter & 1.614 $\pm$ 0.185 & 1.627 $\pm$ 0.183 & 1.620 $\pm$ 0.187 & \underline{1.629 $\pm$ 0.186} & 1.619 $\pm$ 0.187 & \textbf{1.646 $\pm$ 0.187} \\
    \midrule
    \multicolumn{7}{l}{\textbf{Wine Dataset}} \\
    RI & 0.805 $\pm$ 0.117 & 0.824 $\pm$ 0.121 & 0.831 $\pm$ 0.125 & \underline{\textbf{0.865 $\pm$ 0.117}} & 0.815 $\pm$ 0.122 & 0.843 $\pm$ 0.155 \\
    SS & 0.254 $\pm$ 0.047 & 0.263 $\pm$ 0.049 & 0.259 $\pm$ 0.051 & \underline{\textbf{0.271 $\pm$ 0.048}} & 0.256 $\pm$ 0.048 & 0.268 $\pm$ 0.057 \\
    Dist Centroid & 4.144 $\pm$ 0.353 & 4.210 $\pm$ 0.360 & 4.205 $\pm$ 0.362 & \underline{4.300 $\pm$ 0.354} & 4.152 $\pm$ 0.358 & \textbf{4.336 $\pm$ 0.487} \\
    Intra & 45.24 $\pm$ 1.92 & 45.00 $\pm$ 1.98 & 45.00 $\pm$ 2.02 & \underline{\textbf{44.70 $\pm$ 1.92}} & 45.17 $\pm$ 1.96 & 45.11 $\pm$ 2.39 \\
    Inter & 80.36 $\pm$ 3.63 & 81.14 $\pm$ 3.74 & 80.99 $\pm$ 3.84 & \underline{82.13 $\pm$ 3.85} & 80.45 $\pm$ 3.69 & \textbf{82.94 $\pm$ 5.70} \\
    \bottomrule
    \end{tabular}
    \caption{Exact Solutions}
\end{table*}

\begin{table*}[!ht]
    \centering
    \fontsize{9}{11}\selectfont
    \begin{tabular}{lcccccc}
    \toprule
      & Intra & Intra$^*$ & Inter & Intra-Inter combined & Weighted MaxCut & $k$-means \\
    \midrule
    \multicolumn{7}{l}{\textbf{Gaussian Overlapping Dataset}} \\
    RI & 0.767 $\pm$ 0.000 & 0.767 $\pm$ 0.000 & 0.798 $\pm$ 0.000 & \underline{\textbf{0.798 $\pm$ 0.000}} & 0.767 $\pm$ 0.000 & 0.729 $\pm$ 0.000 \\
    SS & 0.433 $\pm$ 0.000 & 0.437 $\pm$ 0.000 & 0.436 $\pm$ 0.000 & \underline{0.444 $\pm$ 0.000} & 0.433 $\pm$ 0.000 & \textbf{0.461 $\pm$ 0.000} \\
    Dist Centroid & 1.960 $\pm$ 0.000 & 1.984 $\pm$ 0.000 & 1.983 $\pm$ 0.000 & \underline{2.012 $\pm$ 0.000} & 1.960 $\pm$ 0.000 & \textbf{2.105 $\pm$ 0.000} \\
    Intra & 128.45 $\pm$ 0.00 & 128.02 $\pm$ 0.00 & 128.12 $\pm$ 0.00 & \underline{127.31 $\pm$ 0.00} & 128.45 $\pm$ 0.00 & \textbf{127.05 $\pm$ 0.00} \\
    Inter & 317.58 $\pm$ 0.00 & 320.12 $\pm$ 0.00 & 319.67 $\pm$ 0.00 & \underline{323.38 $\pm$ 0.00} & 317.58 $\pm$ 0.00 & \textbf{335.91 $\pm$ 0.00} \\
    \midrule
    \multicolumn{7}{l}{\textbf{Iris Dataset}} \\
    RI & 0.922 $\pm$ 0.000 & \underline{\textbf{0.939 $\pm$ 0.006}} & 0.922 $\pm$ 0.000 & 0.922 $\pm$ 0.000 & 0.922 $\pm$ 0.000 & 0.904 $\pm$ 0.000 \\
    SS & 0.434 $\pm$ 0.000 & 0.438 $\pm$ 0.001 & \underline{0.442 $\pm$ 0.000} & \underline{0.442 $\pm$ 0.000} & 0.434 $\pm$ 0.000 & \textbf{0.443 $\pm$ 0.000} \\
    Dist Centroid & 0.096 $\pm$ 0.000 & 0.096 $\pm$ 0.000 & \underline{0.097 $\pm$ 0.000} & \underline{0.097 $\pm$ 0.000} & 0.096 $\pm$ 0.000 & \textbf{0.097 $\pm$ 0.000} \\
    Intra & 4.240 $\pm$ 0.000 & 4.228 $\pm$ 0.003 & \underline{4.219 $\pm$ 0.000} & \underline{4.219 $\pm$ 0.000} & 4.240 $\pm$ 0.000 & \textbf{4.212 $\pm$ 0.000} \\
    Inter & 10.343 $\pm$ 0.000 & 10.378 $\pm$ 0.010 & \underline{10.409 $\pm$ 0.000} & \underline{10.409 $\pm$ 0.000} & 10.343 $\pm$ 0.000 & \textbf{10.429 $\pm$ 0.000} \\
    \midrule
    \multicolumn{7}{l}{\textbf{Wine Dataset}} \\
    RI & 0.817 $\pm$ 0.000 & 0.817 $\pm$ 0.000 & 0.872 $\pm$ 0.004 & \underline{0.888 $\pm$ 0.000} & 0.831 $\pm$ 0.000 & \textbf{0.903 $\pm$ 0.000} \\
    SS & 0.272 $\pm$ 0.000 & 0.272 $\pm$ 0.000 & 0.276 $\pm$ 0.003 & \underline{0.279 $\pm$ 0.000} & 0.273 $\pm$ 0.000 & \textbf{0.280 $\pm$ 0.000} \\
    Dist Centroid & 3.942 $\pm$ 0.000 & 3.942 $\pm$ 0.000 & 3.981 $\pm$ 0.025 & \underline{4.004 $\pm$ 0.000} & 3.950 $\pm$ 0.000 & \textbf{4.013 $\pm$ 0.000} \\
    Intra & 339.96 $\pm$ 0.00 & 339.96 $\pm$ 0.00 & 338.54 $\pm$ 0.80 & \underline{337.69 $\pm$ 0.00} & 339.53 $\pm$ 0.00 & \textbf{337.37 $\pm$ 0.00} \\
    Inter & 580.70 $\pm$ 0.00 & 580.70 $\pm$ 0.00 & 583.75 $\pm$ 1.96 & \underline{585.73 $\pm$ 0.00} & 581.37 $\pm$ 0.00 & \textbf{586.57 $\pm$ 0.00} \\
    \midrule
    \multicolumn{7}{l}{\textbf{0-1 MNIST Dataset}} \\
    RI & 0.948 $\pm$ 0.048 & 0.954 $\pm$ 0.037 & 0.985 $\pm$ 0.013 & \underline{\textbf{0.988 $\pm$ 0.012}} & 0.975 $\pm$ 0.027 & 0.981 $\pm$ 0.015 \\
    SS & 0.472 $\pm$ 0.030 & 0.474 $\pm$ 0.026 & 0.476 $\pm$ 0.022 & \underline{0.479 $\pm$ 0.020} & 0.477 $\pm$ 0.023 & \textbf{0.480 $\pm$ 0.020} \\
    Dist Centroid & 7.659 $\pm$ 0.179 & 7.682 $\pm$ 0.170 & 7.691 $\pm$ 0.133 & \underline{7.717 $\pm$ 0.132} & 7.699 $\pm$ 0.149 & \textbf{7.726 $\pm$ 0.135} \\
    Intra & 986.57 $\pm$ 27.98 & 985.48 $\pm$ 26.67 & 983.61 $\pm$ 23.57 & \underline{\textbf{982.51 $\pm$ 23.67}} & 983.50 $\pm$ 25.05 & 982.64 $\pm$ 23.64 \\
    Inter & 1676.1 $\pm$ 18.55 & 1678.7 $\pm$ 18.73 & 1677.9 $\pm$ 18.93 & \underline{1680.9 $\pm$ 18.94} & 1679.4 $\pm$ 18.92 & \textbf{1682.5 $\pm$ 18.64} \\
    \bottomrule
    \end{tabular}
    \caption{Simulated Annealing Results}
\end{table*}

\begin{table*}[!ht]
    \centering
    \fontsize{9}{11}\selectfont
    \begin{tabular}{lccccc}
    \toprule
      & Intra & Intra$^*$ & Intra-Inter combined & weighted MaxCut & $k$-means \\
    \midrule
    \multicolumn{6}{l}{\textbf{Gaussian Overlapping Dataset}} \\
    RI & 0.497 $\pm$ 0.000 & 0.499 $\pm$ 0.000 & \underline{\textbf{0.798 $\pm$ 0.000}} & 0.497 $\pm$ 0.000 & 0.729 $\pm$ 0.000 \\
    SS & -0.006 $\pm$ 0.000 & -0.004 $\pm$ 0.000 & \underline{0.444 $\pm$ 0.000} & -0.004 $\pm$ 0.000 & \textbf{0.461 $\pm$ 0.000} \\
    Dist Centroid & 0.087 $\pm$ 0.000 & 0.067 $\pm$ 0.000 & \underline{2.012 $\pm$ 0.000} & 0.083 $\pm$ 0.000 & \textbf{2.105 $\pm$ 0.000} \\
    Intra & 189.145 $\pm$ 0.000 & 188.982 $\pm$ 0.000 & \underline{127.307 $\pm$ 0.000} & 189.046 $\pm$ 0.000 & \textbf{127.049 $\pm$ 0.000} \\
    Inter & 189.156 $\pm$ 0.000 & 189.259 $\pm$ 0.000 & \underline{323.381 $\pm$ 0.000} & 189.273 $\pm$ 0.000 & \textbf{335.913 $\pm$ 0.000} \\
    \midrule
    \multicolumn{6}{l}{\textbf{Iris Dataset}} \\
    RI & 0.503 $\pm$ 0.010 & 0.506 $\pm$ 0.013 & \underline{\textbf{0.922 $\pm$ 0.000}} & 0.504 $\pm$ 0.011 & 0.904 $\pm$ 0.000 \\
    SS & 0.005 $\pm$ 0.015 & 0.009 $\pm$ 0.015 & \underline{0.442 $\pm$ 0.000} & 0.005 $\pm$ 0.015 & \textbf{0.443 $\pm$ 0.000} \\
    Dist Centroid & 0.015 $\pm$ 0.007 & 0.017 $\pm$ 0.007 & \underline{0.097 $\pm$ 0.000} & 0.015 $\pm$ 0.007 & \textbf{0.097 $\pm$ 0.000} \\
    Intra & 6.190 $\pm$ 0.059 & 6.176 $\pm$ 0.062 & \underline{4.219 $\pm$ 0.000} & 6.191 $\pm$ 0.059 & \textbf{4.212 $\pm$ 0.000} \\
    Inter & 6.387 $\pm$ 0.106 & 6.412 $\pm$ 0.110 & \underline{10.409 $\pm$ 0.000} & 6.383 $\pm$ 0.107 & \textbf{10.429 $\pm$ 0.000} \\
    \midrule
    \multicolumn{6}{l}{\textbf{Wine Dataset}} \\
    RI & 0.503 $\pm$ 0.008 & 0.503 $\pm$ 0.008 & \underline{0.888 $\pm$ 0.000} & 0.503 $\pm$ 0.009 & \textbf{0.903 $\pm$ 0.000} \\
    SS & 0.003 $\pm$ 0.004 & 0.003 $\pm$ 0.003 & \underline{0.279 $\pm$ 0.000} & 0.003 $\pm$ 0.004 & \textbf{0.280 $\pm$ 0.000} \\
    Dist Centroid & 0.729 $\pm$ 0.153 & 0.749 $\pm$ 0.138 & \underline{4.004 $\pm$ 0.000} & 0.751 $\pm$ 0.149 & \textbf{4.013 $\pm$ 0.000} \\
    Intra & 412.601 $\pm$ 1.056 & 412.484 $\pm$ 0.909 & \underline{337.692 $\pm$ 0.000} & 412.430 $\pm$ 1.006 & \textbf{337.371 $\pm$ 0.000} \\
    Inter & 421.554 $\pm$ 2.453 & 421.830 $\pm$ 2.104 & \underline{585.733 $\pm$ 0.000} & 421.949 $\pm$ 2.388 & \textbf{586.572 $\pm$ 0.000} \\
    \midrule
    \multicolumn{6}{l}{\textbf{0-1 MNIST Dataset}} \\
    RI & 0.501 $\pm$ 0.003 & 0.501 $\pm$ 0.003 & \underline{\textbf{0.989 $\pm$ 0.010}} & 0.501 $\pm$ 0.004 & 0.981 $\pm$ 0.015 \\
    SS & 0.005 $\pm$ 0.007 & 0.005 $\pm$ 0.007 & \underline{0.479 $\pm$ 0.020} & 0.005 $\pm$ 0.007 & \textbf{0.480 $\pm$ 0.020} \\
    Dist Centroid & 1.189 $\pm$ 0.226 & 1.186 $\pm$ 0.216 & \underline{7.717 $\pm$ 0.131} & 1.207 $\pm$ 0.209 & \textbf{7.726 $\pm$ 0.135} \\
    Intra & 1190.06 $\pm$ 25.51 & 1190.11 $\pm$ 25.59 & \underline{\textbf{982.47 $\pm$ 23.59}} & 1189.90 $\pm$ 25.50 & 982.64 $\pm$ 23.64 \\
    Inter & 1207.84 $\pm$ 24.47 & 1207.77 $\pm$ 24.33 & \underline{1680.99 $\pm$ 18.89} & 1208.23 $\pm$ 24.48 & \textbf{1682.49 $\pm$ 18.64} \\
    \bottomrule
    \end{tabular}
    \caption{Quantum Annealing Results}
\end{table*}

\begin{figure*}[!ht]
\begin{minipage}[c]{\linewidth}
    \centering
    \includegraphics[width=0.6\textwidth]{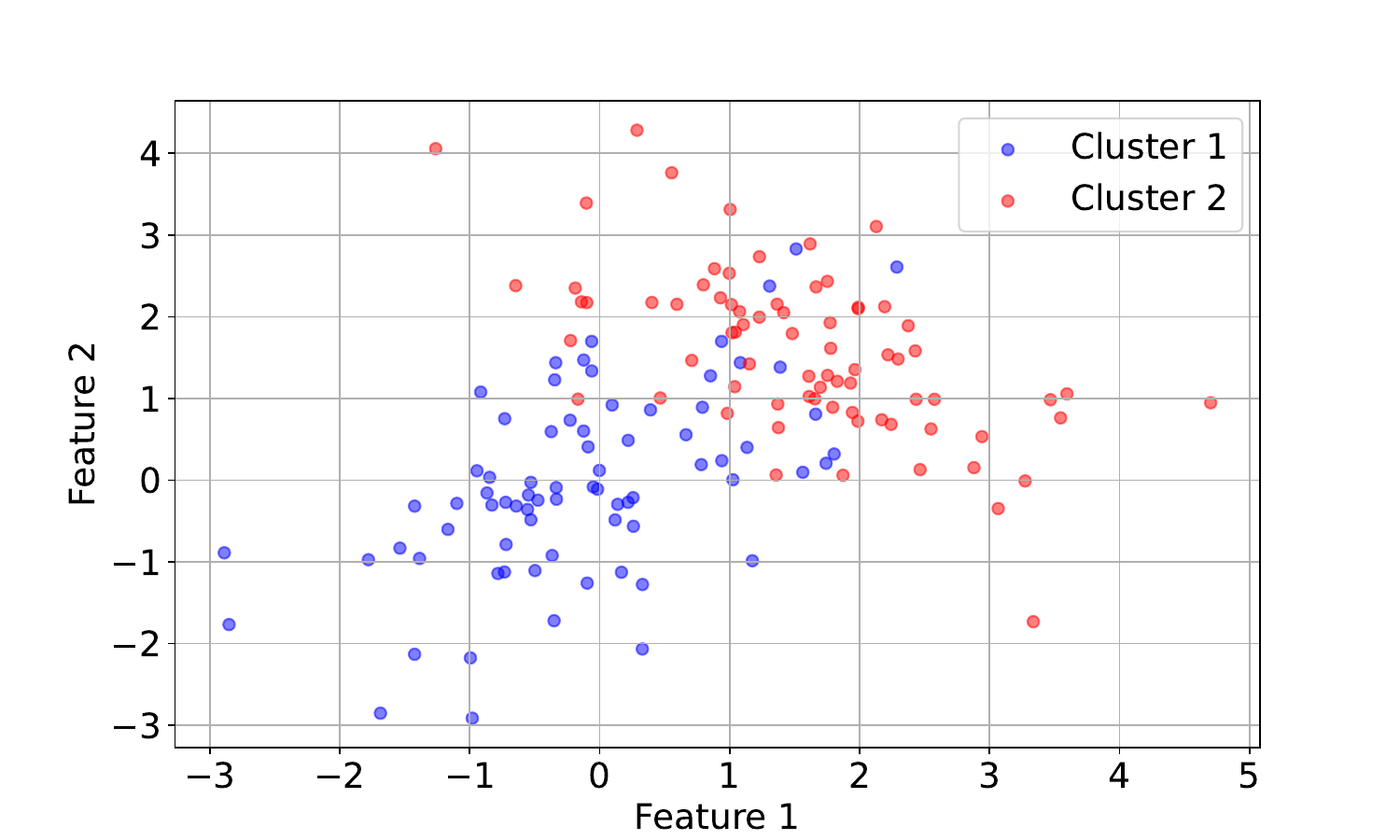}
    \caption{Figure illustrates a 2D Gaussian-distributed synthetic dataset with overlapping clusters, showing Cluster 1 in blue and Cluster 2 in red.}
    \label{fig:gsoverlap}
\end{minipage}
\end{figure*}

\end{document}